\documentclass{aastex62}

\graphicspath{{./}{figures/}}
\received{\today}
\revised{XX-XX-2020}
\accepted{XX-XX-2020}
\submitjournal{ApJ}

\shorttitle{Transit Spectra of Tidally-locked Hot Jupiters}
\shortauthors{Lacy et al.}

\begin{document}

\title{JWST Transit Spectra I: Exploring Potential Biases and Opportunities in Retrievals of Tidally-locked Hot Jupiters with Clouds and Hazes}

\correspondingauthor{Brianna Lacy}
\email{blacy@princeton.astro.edu}

\author[0000-0002-0786-7307]{Brianna I. Lacy}
\affil{Princeton University \\
Peyton Hall, Ivy Lane \\
Princeton, NJ 08544, USA}

\author{Adam Burrows}
\affil{Princeton University \\
Peyton Hall, Ivy Lane \\
Princeton, NJ 08544, USA}

\begin{abstract}
 Many of the exoplanets for which we can obtain the highest $SNR$ transit spectra are tidally locked. The atmospheres on tidally-locked planets likely exhibit large differences between the day and night side of the planet, the poles, and the morning versus evening terminators. However, current state-of-the art retrieval models typically employ only one or two temperature-pressure profiles and corresponding opacity sources, representing the average conditions around the whole limb. In some cases, this will lead to biased measurements, and the approach fundamentally ignores an opportunity to gain empirical information about the 3D structure around the limb of these planets. New methods must be developed as we enter an era of higher-resolution, higher-$SNR$ transit spectra with the broad continuous wavelength coverage expected from upcoming missions like JWST and ARIEL. In this paper, we illustrate how the combined effects of aerosols and day-night temperature gradients shape transit spectra of tidally-locked exoplanets when full 3D structures are taken into account and evaluate the implications for retrievals of atmospheric properties. To do this, we have developed a new code, METIS, which can compute transit spectra for an arbitrary longitude-latitude-altitude grid of temperatures and pressures. Using METIS, we pair flexible treatments of clouds and hazes with simple parameterized day-night temperature gradients to compute transit spectra and perform retrieval experiments across a wide array of possible exoplanet atmospheric properties. Our key findings are that: (1) the presence of aerosols typically increases the effects of day-night temperature gradients on transit spectra; (2) ignoring day-night temperature gradients when attempting to perform Bayesian parameter estimation will still return biased results when aerosols are present, as has already been shown for clear atmospheres in the literature; (3) when a day-night temperature gradient is present and accounted for in the retrieval model, some transit spectra can provide sufficient information to constrain temperatures and the width of the transition from day to night. The presence of clouds and hazes can actually tighten such constraints, but also weaken constraints on metallicity. This paper represents a step towards the larger goal of developing models and theory of adequate complexity to match the superior quality data that will soon be available.

 \end{abstract}

\section{Introduction} \label{sec:intro}

Studying exoplanet atmospheres is compelling for a number of reasons. Among these are the hope to uncover trends related to planet formation and evolution (\citealt{Oberg2011}; \citealt{Piso2016}), make comparisons to Solar System planets (\citealt{Seager2010}; \citealt{Madhusudhan2019}), understand the climate and conditions on alien worlds (\citealt{Heng2015}; \citealt{Burrows2014}), and even search for life beyond Earth (\citealt{Meadows2018}; \citealt{Kopparapu2019}). One technique for studying exoplanet atmospheres is to measure the slight variation in transit depth with radius (\citealt{Seager2000}; \citealt{Brown2001}; \citealt{Hubbard2001}; \citealt{Charbonneau2002}; \citealt{Burrows2006a}). At wavelengths of light where molecules in the atmosphere have large opacities, the planet will block out more light from the host star compared to wavelengths of light where the atmosphere is more transparent. Transit spectroscopy has already been used to detect molecules and atomic species in exoplanet atmospheres (\citealt{Deming2013}; \citealt{Spake2018}), infer the presence of clouds and hazes (\citealt{Barstow2013}; \citealt{Fraine2013}; \citealt{Crossfield2013}; \citealt{Kreidberg2014b}; \citealt{knutson2014a}; \citealt{knutson2014b}; \citealt{Sing2016}; \citealt{Louden2017}; \citealt{Wakeford2019}), and make measurements of water abundances (\citealt{Fraine2014}; \citealt{Kreidberg2014a}; \citealt{Barstow2017}; \citealt{Wakeford2018}; \citealt{Pinhas2019}). 

Future missions like the James Webb Space Telescope (JWST) and the Atmospheric Remote-sensing Infrared Exoplanet Large-survey (ARIEL) will enable transit spectra with high resolution (R$>$100) over a broad continuous wavelength range extending further into the infrared (IR) than currently possible (\citealt{Greene2016}; \citealt{Stevenson2016}; \citealt{Puig2016}; \citealt{Tinetti2018}; \citealt{Zellem2019}). It is hoped that JWST and ARIEL will achieve more precise measurements of water abundances, measure the first abundances of other species like CH$_4$ and CO, combine these to estimate C/O ratios and metallicities, identify aerosol species, obtain spectra of smaller planets, and recognize signatures of inhomogeneities between morning/evening  terminators (\citealt{Greene2016}; \citealt{Wakeford2017}; \citealt{Bean2018};  \citealt{Powell2019}; \citealt{Zellem2019}). With JWST, we expect to learn about tens of planets in exquisite detail. ARIEL will observe with narrower wavelength coverage and lower signal-to-noise ratio ($SNR$), but it will target 500-1000 planets, allowing researchers to look for overarching patterns and trends. 

The anticipation of improved data quality and quantity warrants careful examination of current retrieval techniques used to perform Bayesian parameter estimation on transit spectra. Often retrievals using transit spectra are done with one parameterized temperature-pressure (T-P) profile, parameterized aerosol opacities, and free parameters for the abundances of major gaseous absorbers such as H$_2$O, CO, HCN, CO$_2$ and CH$_4$. \citealt{Barstow2020a} and \citealt{Barstow2020b} provide a recent review and comparison of existing retrieval codes\footnote{examples of transit retrieval codes which allow aerosols include: NEMESIS (\citealt{MacDonald2017}), PyRat-Bay (\citealt{Cubillos2017}), BART (\citealt{Blecic2017}), SCARLET (\citealt{Fraine2014}), CHIMERA (\citealt{Line2013}), $\tau$-REx (\citealt{Waldmann2015})}. In some cases, a 1D model is perfectly adequate, but, in some cases, it has been shown to fall short, such as for exoplanets with patchy clouds (\citealt{Line2016}; \citealt{MacDonald2017}) or exoplanets with significant differences between the morning and evening terminator (\citealt{Powell2019}; \citealt{Kempton2017a}; \citealt{MacDonald2020}), and exoplanets with significant differences in temperature and chemistry between the day-side and night-side (\citealt{Caldas2019}; \citealt{Pluriel2020}). Using a combination of two 1D profiles has already been shown to mitigate some of these biases (e.g. \citealt{MacDonald2017}; \citealt{MacDonald2020}), but more work is needed to ascertain exactly when more complex models are applicable and how best to parameterize the expected physical conditions of exoplanet atmospheres. 

One may be tempted to simply select targets for JWST and ARIEL which seem less susceptible to the complexities arising from inhomogeneous atmospheres. However, many of the strongest targets for transit spectroscopy will be tidally locked due to observational biases. Warm, H$_2$-He dominated atmospheres with large radii and large scale heights yield the highest $SNR$ for transit spectroscopy observations. Furthermore, short periods allow for easy scheduling of observations and stacking many separate transit observations can reduce noise further. Objects which meet the above criteria tend to orbit close to their host star, which results in synchronous rotation on relatively short timescales (\citealt{Goldreich1966}; \citealt{Bodenheimer2001}). Hot Jupiters also allow additional characterization of their atmospheres by other methods such as phase curves and emission spectroscopy. Theoretical models which can consistently replicate observations with all three methods will provide the most comprehensive and convincing picture of an alien climate (\citealt{Heng2015}). Terrestrial planets in the habitable zones of M dwarfs are also likely to be tidally locked \citep{Barnes2017}, and such systems currently provide the most promising option for studying the atmospheres of temperate earth-like worlds with transit spectroscopy. It is thus well worth the effort to model and understand the transit spectra of tidally-locked exoplanets. 

We present a new code for modeling transit spectra: Multi-dimensional Exoplanet TransIt Spectra (METIS). METIS takes an arbitrary longitude-latitude-altitude grid of temperatures and pressures along with an overall atmospheric metallicity as input, and then computes the corresponding transit spectra. It assumes hydrostatic equilibrium to map pressures to radii, an ideal gas equation of state, single scattering, and thermochemical equilibrium with rain-out to infer abundances. We use pre-mixed opacity tables from \citealt{Sharp2007}, which span metallicities between 0.1 and 3.16 times solar assuming a solar C/O ratio. The user can specify a variety of clouds and hazes to incorporate. Aerosols in this work are treated as Mie-scattering homogeneous spheres. Provided a parameterized 1, 2 or 3D temperature-pressure structure, METIS can also be used for Bayesian parameter estimation. 

In this work, we focus specifically on hot Jupiters with day-night temperature gradients and inhomogeneous aerosol coverage that varies between the day and night side of the planet, but METIS is well-suited to computing transit spectra for atmospheres which vary between morning, evening and poles as well. We use METIS to carry out Markov Chain Monte-Carlo (MCMC) retrievals and conduct studies of parameter sensitivity in order to investigate the following questions: 
\begin{enumerate}
    \item Which planets (out of those with measured day and night temperatures) are likely to have detectable effects from day-night temperature gradients in their JWST transit spectra? 
    \item Can accurate planet properties be retrieved from JWST transit spectra of tidally-locked hot Jupiters? 
    \item Do aerosols make it more or less important to account for the presence of a day-night temperature gradient in potential targets?  
    \item What are the biases on retrieved metallicities and temperatures from ignoring day-night temperature gradients for clear, hazy, and cloudy atmospheres?
\end{enumerate}

The remainder of this paper is organized as follows: section \ref{sec:context} places our study in the context of previous relevant theoretical studies and observations of tidally-locked hot Jupiters, \S \ref{sec:methods} describes METIS, our newly developed code for computing transit spectra from 3D atmospheres, \S \ref{sec:clear_paramstudy} uses models of clear transit spectra for 12 objects with phase-curve estimates of day-side and night-side temperatures to demonstrate the effects of day-night temperature gradients across different temperatures and surface gravities, \S \ref{sec:aerosol_paramstudies} explores the combined effects of aerosols and day-night temperature gradients through representative parameter sensitivity studies, \S \ref{sec:mcmc_results} presents the results from a series of retrieval experiments on clear, cloudy, and hazy atmospheres. Finally, we summarize and draw our conclusions in \S \ref{sec:conclusions}. We also provide an appendix with auxiliary descriptions of our models and methods.

\subsection{Further Context}\label{sec:context}

Tidally-locked hot Jupiters are the most thoroughly modelled and observed class of exoplanet to date (\citealt{Heng2015}). Their large radii, short periods, and high temperatures allow one to obtain infrared phase curves, emission spectra and transit spectra with relatively high $SNR$'s compared to other types of exoplanets (eg: \citealt{Charbonneau2000}; \citealt{Henry2000}; \citealt{Cowan2007}; \citealt{Cowan2012}; \citealt{Knutson2007}; \citealt{Knutson2009a}; \citealt{Knutson2009b}, \citealt{Knutson2012}; \citealt{Borucki2009}; \citealt{Demory2013}; \citealt{Maxted2013}; \citealt{Stevenson2014}; \citealt{Zellem2014}; \citealt{Wong2015b}). Due to the extreme irradiation directed at the permanent day-side of these planets, it is theorized that their atmospheres will have large differences in temperature between the day side and the night side, and that extremely fast-flowing equatorial jets will emerge (\citealt{Showman2002}; \citealt{Cooper2005}; \citealt{Fortney2008}; \citealt{Showman2008b}; \citealt{Lewis2010}; \citealt{Rauscher2010}; \citealt{Cowan2011}; \citealt{Menou2012}; \citealt{Perna2012}; \citealt{Perez-Becker2013}; \citealt{Heng2015}; \citealt{Kataria2016}; \citealt{Ginzburg2016}; \citealt{Komacek2016}). These winds should also cause an east-west asymmetry in addition to the day-night differences in the atmospheric properties (\citealt{Kempton2017a}; \citealt{Powell2019}). 
 
Observations of hot Jupiters have demonstrated a diverse range of behaviors, and theory can explain some, but not all, of what is going on. For example, phase curves have shown that the fractional difference between day-side and night-side emission and the degree of hot-spot offset can vary significantly from planet to planet (\citealt{Komacek2016}; \citealt{Keating2019}). Earlier observations indicated an increasing trend in day-night temperature contrast with level of stellar irradiation (\citealt{Komacek2016}). Theoretical modeling with Global Circulation Models (GCM) could explain this trend, at least qualitatively, as the result of the competition between the timescale for advection against the timescale for radiative cooling and heating (\citealt{Komacek2016}). A recent reanalysis by \citealt{Keating2019} indicates that night-side temperatures of hot Jupiters seem to clump around 1100 K with a slight upward trend for only the most highly irradiated objects. The favored explanation for this is that night-side clouds with similar properties on all these objects re-radiate heat at around 1100 K. Regardless of the precise mechanisms underlying the cause of day-night temperature gradients, they are almost certainly present. Linking the interpretation of transit spectra to phase curve observations by attempting to extract multi-dimensional information, could provide a more complete picture of the physical conditions on these distant bodies (\citealt{Heng2015}; \citealt{Lee2019}). 

Considering both the observational evidence, and theoretical models, it seems safe to say that the limbs of hot Jupiters probed by transit spectra will be very heterogeneous. Many researchers have noted this, and investigated how the full 3D properties of hot Jupiters as modelled by GCMs may manifest in transit spectra. These studies generally pull out the atmosphere's structure in the annulus right around the terminator of a converged GCM, then use some physically motivated assumptions to post-facto fill in chemistry and aerosols, and finally compute transit spectra (\citealt{Burrows2010}; \citealt{Fortney2010}; \citealt{Dobbs-Dixon2012}; \citealt{Parmentier2013}; \citealt{Charnay2015a}; \citealt{Charnay2015b}; \citealt{Lee2015}; \citealt{Helling2016}; \citealt{Lee2016}; \citealt{Line2016};
\citealt{Kempton2017a}; \citealt{Lines2018}; \citealt{Caldas2019}; \citealt{Helling2019b}; \citealt{Powell2019}; \citealt{Lee2019}; \citealt{Pluriel2020}; \citealt{MacDonald2020}; \citealt{Taylor2020}). Comparing these 3D transit spectra to 1D transit spectra has revealed detectable differences. When theoretical 3D spectra are fed into 1D retrieval frameworks and compared to ground truth parameters, it is evident that current state of the art approaches will lead to biases once we have higher quality data. Most studies of inhomogeneous limbs have focused on variation about the azimuthal extent of the terminator, not differences along the transverse path. But the assumption of homogeneity along the transverse path does not always hold, and it is most likely to fail for tidally-locked hot Jupiters \citep{Caldas2019}.

\citealt{Caldas2019} point out that transit spectra of hot puffy atmospheres probe a relatively broad swath of atmosphere about the terminator, and, for tidally-locked exoplanets, this region will likely vary widely in temperature and chemical composition between the day and night side (\citealt{Pluriel2020}). They demonstrated that day-night differences will affect transit spectra at detectable levels and conducted a series of retrieval experiments to determine how this might bias 1D retrievals. To generate input data for the retrieval experiments, they constructed idealized 3D atmosphere structures which varied continuously from a high temperature on the day side to a cooler temperature on the night side. They simulated a wide range of temperatures and transition widths, then attempted to fit them with single T-P profile transit spectra as is standard practice for retrievals. They found that they could produce satisfactory fits to the simulated data based on goodness-of-fit metrics, but got results which systematically biased the terminator temperature up towards the day-side temperature. This warmer temperature then biased retrieved water abundances. They thus concluded that ignoring the presence of day-night temperature gradients will confuse our understanding of some of the strongest observational targets and potentially hinder attempts to understand overarching trends in large samples of exoplanet atmospheres.

In this paper we extend \citealt{Caldas2019}'s line of investigation further by modeling day-night temperature gradients with local thermal-equilibrium chemistry. This approach varies the abundances of all major opacity sources continuously across the day-night transition, whereas their study kept the chemistry constant through out the whole atmosphere, and a follow up study (\citealt{Pluriel2020}) varied only the H$^{-}$ abundance between the day and night sides of the planet. We also explore the effects of including opacity from a variety of clouds and hazes. Including aerosols is an important step because existing transit spectra directly show that they are present at the limbs of some tidally-locked exoplanets. There are also hints that reflective aerosols may be present on the day-side in some hot Jupiter atmospheres arose when Kepler provided the first observations of optical phase curves \citep{Parmentier2016}. Other observations have indicated that different species may be condensing on the cooler night-side then evaporating on the warmer day side (\citealt{Keating2019}; \citealt{Ehrenreich2020}). \citealt{Gao2020} expect silicate aerosols to form in atmospheres with equilibrium temperatures between 1000 and 2000 K, and hyrdrocarbon hazes to form for equilibrium temperatures below 1000 K, based upon the results of detailed microphysical models. 

In \S \ref{sec:mcmc_results} we show the results of similar retrieval experiments to those done by \citealt{Caldas2019}, where data is simulated for an atmosphere with a day-night temperature gradient but then fit using a single T-P profile. We also show results for retrievals done with the full temperature gradient model, in order to explore the issue from a more positive perspective. If ignored and brushed under the rug, day-night temperature gradients will most likely bias temperature and abundance measurements, but, if acknowledged and included in models, day-night temperature gradients may provide additional information about the extreme and dynamic atmospheres of highly-irradiated planets. Before we show the retrieval results, we illustrate the nature of our temperature gradient model for clear atmospheres in \S \ref{sec:clear_paramstudy} and for hazy or cloudy atmospheres in \S \ref{sec:aerosol_paramstudies}.

\section{Methods} \label{sec:methods}
\subsection{Computing Transit Spectra}
METIS produces transit spectra given an atmospheric metallicity and an arbitrary longitude-latitude-altitude grid of temperatures, pressures, and densities (the type of output expected from a General Circulation Model, GCM). METIS is built around our own implementation of the geometrical framework and algorithms of Pytmosph3R \cite{Caldas2019}. We used the detailed descriptions in \cite{Caldas2019} to replicate the basic framework of their code, but then paired the geometry and integration approaches with a suite of pre-tabulated equilibrium chemistry and corresponding gas opacities rather than using their on-the-fly approach to chemistry and opacity calculations.

The transit spectrum for a 3D atmosphere can be computed as:
\begin{equation}\label{eq:transit_depth}
   \frac{\delta F}{F}\big(\lambda\big) = \frac{\int_{0}^{2\pi}\int_{0}^{\infty}b(1-e^{-\tau(\theta,b,\lambda)}) dbd\theta}{2\pi R_*(\lambda)^2} ~~,
\end{equation}
 where $\frac{\delta F}{F}\big(\lambda\big)$ refers to the fractional change in flux during mid-transit or the so-called transit depth, $b$ is the impact parameter of a ray of light passing through the planet's atmosphere relative to the center of the planet, $\theta$ is the azimuthal angle about a line drawn from the star through the center of the planet to the observer, $\tau(\theta,b,\lambda)$ is the slant optical depth of the atmosphere at the specified $\theta$ and $b$, and $R_{*}$ is the radius of the host star.  We construct $\tau(\theta,b,\lambda)$ as:
 \begin{equation}\label{eq:slant_opt_depth}
     \tau(\theta,b,\lambda) = \int_{-\infty}^{+\infty} \rho(\theta,b,x)\kappa(\theta,b,x,\lambda) dx ~,
 \end{equation}
where $\rho$ is the mass density of the atmosphere, $\kappa$ is the opacity expressed as a cross section over a mass, and $x$ is path length along the light ray described by cylindrical coordinates $b$, $\theta$. Note that $x$ and $b$ can be related to an altitude in the atmosphere $z$ by the Pythagorean theorem: $z = \sqrt{b^2+\mid x\mid^2} - R_P$, where we have set $x$=0 at the terminator and the altitude $z$ is measured relative to a planet radius $R_P$.

In practice we discretize the integral in equation (\ref{eq:slant_opt_depth}) as a summation over all the atmosphere grid cells passed through by each ray defined by $\theta$ and $b$, varying the path length $dx$ appropriately. We also discretize the integral in equation (\ref{eq:transit_depth}), selecting a lower bound for the impact parameter $b$ that lies below where the planet appears opaque and an upper bound that is sufficiently high for atmospheric extinction to become negligible at all wavelengths. The linear spacings in b are chosen such that resultant transit spectra no longer changed beyond machine precision. This typically takes of order 100 impact parameter bins.

\subsection{Simplified Atmospheric Structures}\label{sec:tgrad_structure}

Throughout our study, rather than run tens of thousands of full GCMs, we construct 3D atmospheres with parameterized day-night temperature gradients assuming hydrostatic equilibrium in the vertical direction, thermo-chemical equilibrium, and the ideal gas law. These assumptions are again inspired by the work of \cite{Caldas2019}. This structure model uses eight parameters to determine a latitude-longitude-altitude grid of temperatures, pressures and densities: the day-side temperature, T$_{day}$, the night-side temperature, T$_{night}$, the angle over which the transition from day-side to night-side temperature occurs, $\beta$, the mass of the planet, M$_p$, the metallicity of the planet's atmosphere, $Z$, the radius of the planet, R$_0$, and finally the reference pressure, P$_0$, to which R$_0$ corresponds. To be physically motivated in using a single value of R$_0$ and P$_0$, they ought to be chosen such that they are deep enough in the atmosphere that the day-night temperature difference imposed by stellar irradiation has faded and the internal state of the planet is dominating (below 10 bars, \citealt{Heng2015}).

The temperatures are mapped onto latitude assuming a linear change from night-side temperature to day-side temperature across an angular width $\beta$: 
\[  T(\phi) = \left\{
\begin{array}{ll}
      T_{night} & \phi\geq \pi/2-\beta/2 \\
      \frac{T_{day}-T_{night}}{\beta}\phi + \frac{T_{day}+T_{night}}{2} & -\beta/2 < \phi < \beta/2\\
      T_{day} & \phi \leq \pi/2+\beta/2~.\\
\end{array} 
\right .\]

Note that here a latitude of 90$^{\circ}$ or $\pi/2$ radians corresponds to the terminator, a latitude of 180$^{\circ}$ or $\pi$ radians corresponds to the substellar point, and a latitude of 0$^{\circ}$ or 0 radians corresponds to the anti-stellar point. This is flipped from the usual orientations of latitude and longitude because it allows us to easily introduce azimuthal symmetry along the axis defined by star-planet-observer. This azimuthal symmetry allows us to speed up the computation of the transit spectrum by removing the integration over $\theta$ in eq. (\ref{eq:transit_depth}). The day-night variation comes in as one integrates along $dx$ in eq. (\ref{eq:slant_opt_depth}), which is now independent of $\theta$. This speed-up allows us to carry out retrievals. 

For each temperature we then assign appropriate pressures to a grid of altitudes according to an isothermal atmosphere with varying gravity: 
\begin{equation}\label{eq:tgrad}
    ln\Big(\frac{P(z_{i+1})}{P(z_i)}\Big)=-\frac{\mu g(z_i)}{k_B T}\Big(\frac{z_{i+1}-z_{i}}{1+\frac{z_{i+1}-z_{i}}{R_p + z_i}}\Big) ~,
\end{equation} where z is 0 at $R_P$, and increases with decreasing pressure. We begin at $P_0$ and z=0 assuming g(0)=$\frac{GM_P}{R_P^2}$, and then integrate upwards from there. Each new pressure level gives a new density following the ideal gas law, and adding on the corresponding mass for that shell and accounting for the new radius gives a new surface gravity. We also allow the mean molecular weight to vary with altitude as the composition of the atmosphere changes, using our pre-tabulated equilibrium chemistry tables to determine the appropriate mean molecular weight.

\begin{figure}
\includegraphics[width=0.288\textwidth]{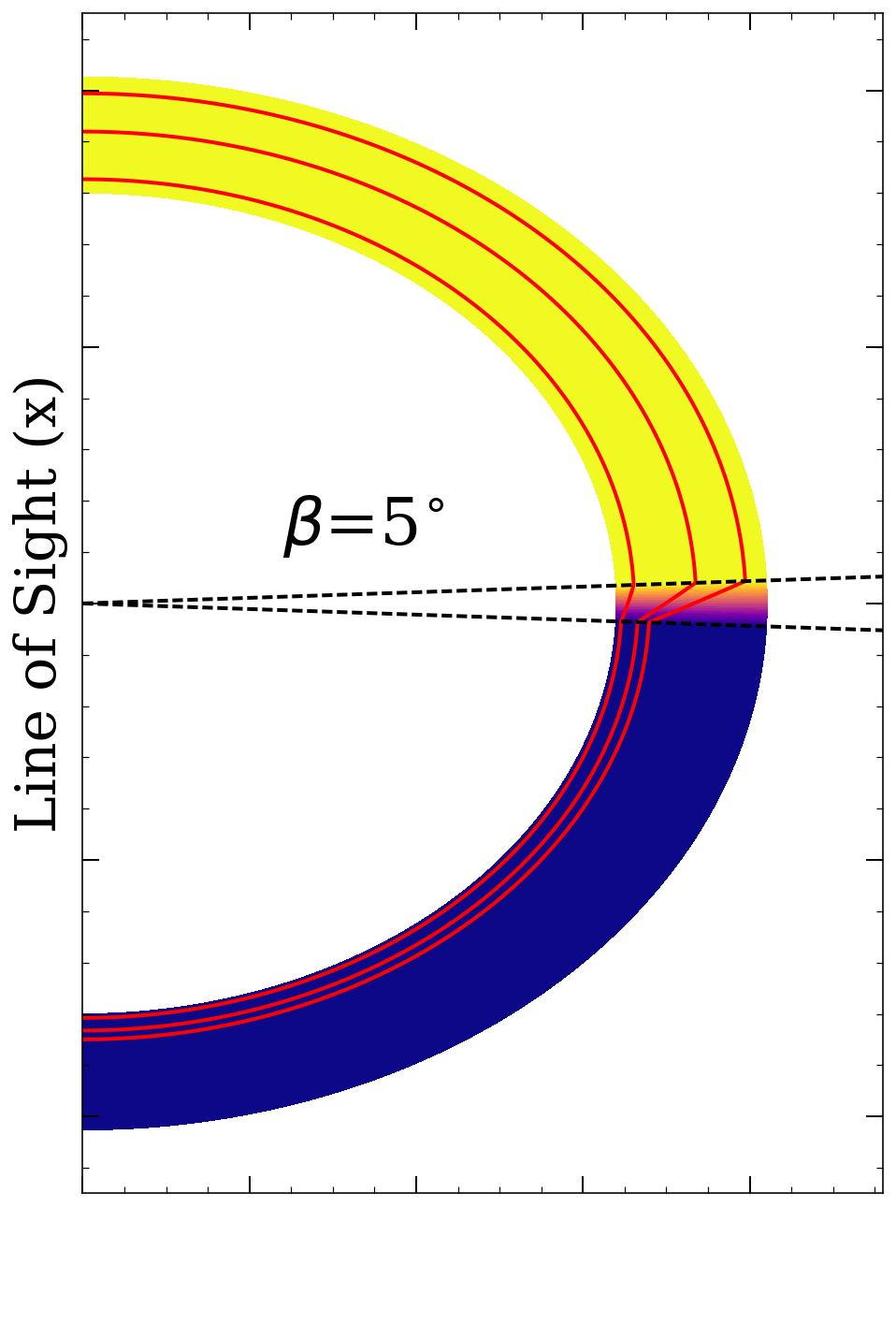}
\includegraphics[width=0.288\textwidth]{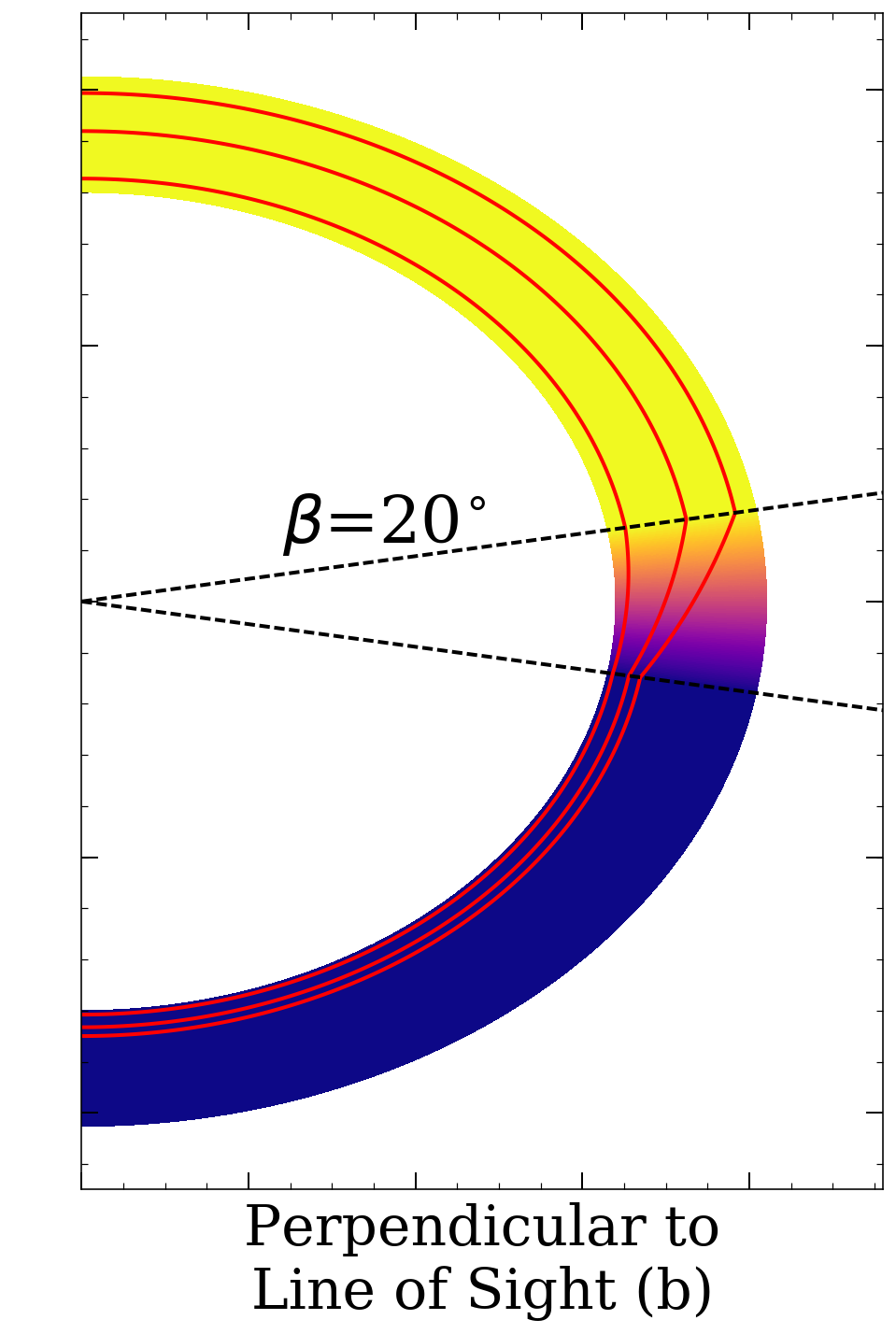}
\includegraphics[width=0.38\textwidth]{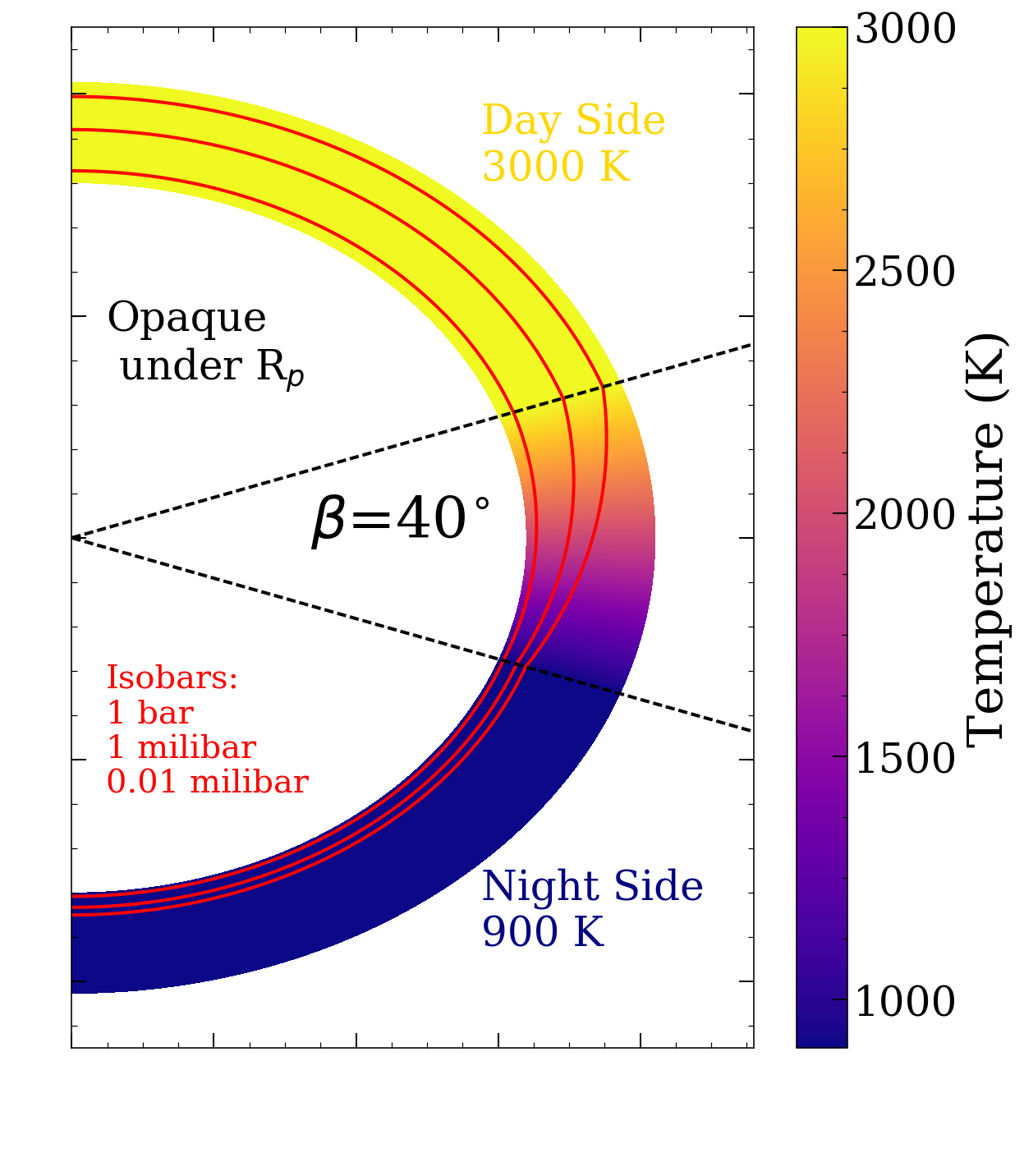}
\caption{Demonstration of the simplified atmospheric structures employed throughout this work. It has a day-night temperature gradient parameterized by a day-side temperature, a night-side temperature, and an angular width ($\beta$) over which the transition from one to the other occurs. At any given latitude-longitude point, the T-P profile is isothermal with the appropriate temperature for that longitude. For this figure, we use an exaggerated contrast of 900 K for the night side and 3000 K for the day side, and a small planet mass so that there is a large scale height: 0.5 M$_J$ with R$_0$=1 R$_J$ and P$_0$=1 bar. $\beta$ varies from a very narrow 5$^{\circ}$ on the left, to roughly what is expected from GCMs in the center (20$^{\circ}$), to a more gradual 40$^{\circ}$ on the right.  \label{fig:toymodel}}
\end{figure}

In reality, transit spectra probe enough of a range in altitude that an isothermal assumption is not fully adequate \citep{Blecic2017}, but this model will still enable us to quantify the effects of having a day-night temperature gradient by representing the limit where all temperature variations are across longitude rather than altitude. An exaggerated example with a 3000 K day-side and 900 K night-side is shown in Figure \ref{fig:toymodel} for three different values of $\beta$. In this figure, the host star would lie towards the top of the Figure and the observer towards the bottom. The impact parameter $b$ would vary along the  horizontal axis, and the path along the line-of-sight would run parallel to the vertical axis.

Once the initial atmospheric structure is set up, either from a GCM output, or from our toy model, we can use the temperatures, pressures, and densities in each cell to determine the chemistry and opacities as described in the following \S \ref{sec:chemistry} and \S \ref{sec:opacities}, then carry out the integrals in eqs. (\ref{eq:slant_opt_depth}) and (\ref{eq:transit_depth}).

In cases where we do not wish to consider a day-night temperature gradient, we can simply assign the same temperature regardless of longitude, and proceed with the same calculations. We refer to such atmospheres as ``uniform" from here on out.

\subsection{Chemistry}\label{sec:chemistry}
We assign mixing ratios for important species using pre-tabulated Tables of equilibrium chemistry computed by \citealt{Burrows1999} and \citealt{Sharp2007}. The calculations used to construct these tables start with the solar composition of 27 chemical elements scaled depending on the metallicity of the table (solar abundances taken from \citealt{Anders1989}). Then a Newton-Raphson solver is used to minimize the total free energy of the system tracking 330 gas-phase species, including the monatomic forms of the elements, as well as about 120 condensates. The calculations also include electrons, H$^-$, and H$^+$. Refractory elements (Ti, V, H$_2$O, and Fe) are withdrawn from the equilibrium amounts in stoichiometric ratios via an ad hoc algorithm to approximate condensate rainout. For metallicities other than those for which we have Tables, we interpolate linearly in log(Z/Z$_{\odot}$). 
 
The tables record the mixing ratios of 30 species for a grid of 805 temperatures evenly spaced in log between 50 K and 5000 K, and 108 pressures evenly spaced in log between 400 atm and 7.982$\times 10^{-9}$ atm.

When aerosols are included, the material tied up in the aerosols is not taken into account in the calculations of gas phase chemistry, so we are essentially assuming that the time scales for photochemical processes and condensation are long compared to gas-phase interactions. We also assume that replenishment of new material from deeper in the atmosphere keeps the gas phase unchanged, or that the amount of material tied up in aerosols is negligible compared to the gas-phase abundances of relevant atomic species.

\subsection{Opacity Sources}\label{sec:opacities}

\subsubsection{Gaseous Opacities}
We use the gaseous opacities from pre-tabulated, pre-mixed tables as described in \cite{Sharp2007}, with updated CH$_4$ opacities from \cite{Yurchenko2014}. Opacities for 26 different ionic, atomic, and molecular species are calculated line-by-line then combined to obtain total opacities according to the chemical equilibrium Tables described in \S\ref{sec:chemistry}. For most species, line-profiles come from collisions with a H$_2$-He atmosphere, but for Na and K, special care is taken. Opacities from H$_2$-H$_2$ and H$_2$-He collision induced absorption are also incorporated into this total opacity. The tables have 5000 spectral bins from 0.3 to 300 $\mu$m evenly spaced in log frequency, 50 temperatures ranging from 50 K to 5000 K evenly spaced in log, and 50 densities ranging from 0.01 to 10$^{-12}$ g/cm$^{3}$ evenly spaced in log. Rayleigh scattering cross sections for the appropriate mixture of gases are provided in a separate Table at the same temperatures and densities but only for a single reference wavelength of $\lambda_0=$1$\mu$m. This cross section is then scaled as $(\lambda/\lambda_0)^{-4}$. 

\subsubsection{Aerosol Opacities}\label{sec:aerosols}
An aerosol is a solid or liquid particle suspended in a gas. Note that throughout this work we use the word \textit{haze} to refer to aerosols formed via photochemistry, the word \textit{cloud} to refer to aerosols that condensed, and the more general term \textit{aerosol} when a statement applies to both hazes and clouds. Haze particles tend to be smaller than cloud particles, though they can actually grow quite large in some circumstances. Incorporating aerosol opacities into a transit model requires extinction cross sections for a range of particle sizes and wavelengths of light, and the number of particles of each size present in the atmosphere at any given location. For the first step, calculating cross sections, we use Mie theory. Mie theory solves Maxwell's equations for light passing through a homogeneous sphere. It says that the scattering efficiency factor and extinction efficiency factor are given by:
\begin{equation}\label{eq:Qsca}
    Q_{sca}(a,x) \equiv \frac{\sigma_{sca}}{\pi a^2} = \frac{2}{x^2}\sum_{n=1}^{\infty} (2n+1)(\mid a_n \mid ^2 + \mid b_n \mid ^2)
\end{equation}
 and 
\begin{equation}\label{eq:Qext}
    Q_{ext}(a,x) \equiv \frac{\sigma_{ext}}{\pi a^2} = \frac{2}{x^2}\sum_{n=1}^{\infty} (2n+1)Re(a_n + b_n)~,
\end{equation}
where $\sigma_{sca}$ is the scattering cross section, $\sigma_{ext}$ is the extinction cross section, $a$ is the particle radius, $x$ is the size parameter 2$\pi a/\lambda$, and $a_n$ and $b_n$ are the Mie coefficients. These coefficients can be expressed in terms of the complex index of refraction of the aerosol material $m(\lambda) = n(\lambda) - ik(\lambda)$, and Bessel Functions of fractional order. 

We used the package {\tt PyMieScat}\footnote{https://pymiescatt.readthedocs.io/en/latest/} to prepare pre-tabulated grids of the cross sections for 5000 wavelengths ranging from 0.4 to 25 $\mu$m and  particle sizes in 60 bins with logarithmic spacing between 0.001 and 100 $\mu$m. These tables are then used to compute the total aerosol opacity for a chosen particle size distribution on the fly in different transit spectra. Most indices of refraction were taken from the compilation done by \citealt{Kitzmann2018}. 

The second aspect of incorporating aerosol opacities is to specify how many particles of a given size are present at any given place in the atmosphere and sum over them. Any type of parameterized aerosol opacity which specifies a total cross section in terms of wavelength, temperature, and pressure could be easily incorporated into METIS. \citealt{Mai2019} and \citealt{Barstow2020c} provide excellent summaries of the usual treatments adopted for aerosols in transit spectroscopy retrievals. These range from the fully data-driven gray absorber + Rayleigh slope model, to the more microphysically-motivated \citealt{Ackerman2001} formulation balancing turbulent diffusion and gravitational settling. Another often adopted practice for condensing species is a phase equilibrium cloud \citep{Charnay2018}. This approach places cloud bases at the intersection of the Claussius-Clapeyron line and the temperature-pressure profile of the atmosphere, and then incorporates only the material which is in excess of the saturation vapor pressure into particles. It does not provide any guidance on the nature of the particle size distributions.

\begin{table}[]\label{tab:aerosol_spatial_parameters}
    \centering
    \begin{tabular}{l|c|l|l}
       Name  & Parameters & Meaning & Intended Aerosol Type \\
    \hline
        Slab &P$_{top}$ & Top pressure cut-off & Condensing clouds \\
        (Fig. \ref{fig:aerosol_spatial_dists}, right)&F & Fraction of available material&or photochemical hazes\\
        &&that contributes to aerosol &\\
        &a$_m$& Modal particle radius&\\
        &$\sigma _a$& Size dispersion for log-normal &\\
    \hline
        Phase Equilibrium &$\alpha$& Ratio of gas scale height to &Condensing clouds\\
        (Fig. \ref{fig:aerosol_spatial_dists}, left)&&aerosol scale height&\\
        &a$_m$& Modal particle radius&\\
        &$\sigma _a$& Size dispersion for log-normal &\\
    \end{tabular}
    \caption{Summary of aerosol spatial parameterizations. These can be paired with other size distributions besides the log normal parameters included here. Options to use a wide variety of aerosol species are available.}
\end{table}

\begin{figure}
    \centering
    \includegraphics[width=0.475\textwidth]{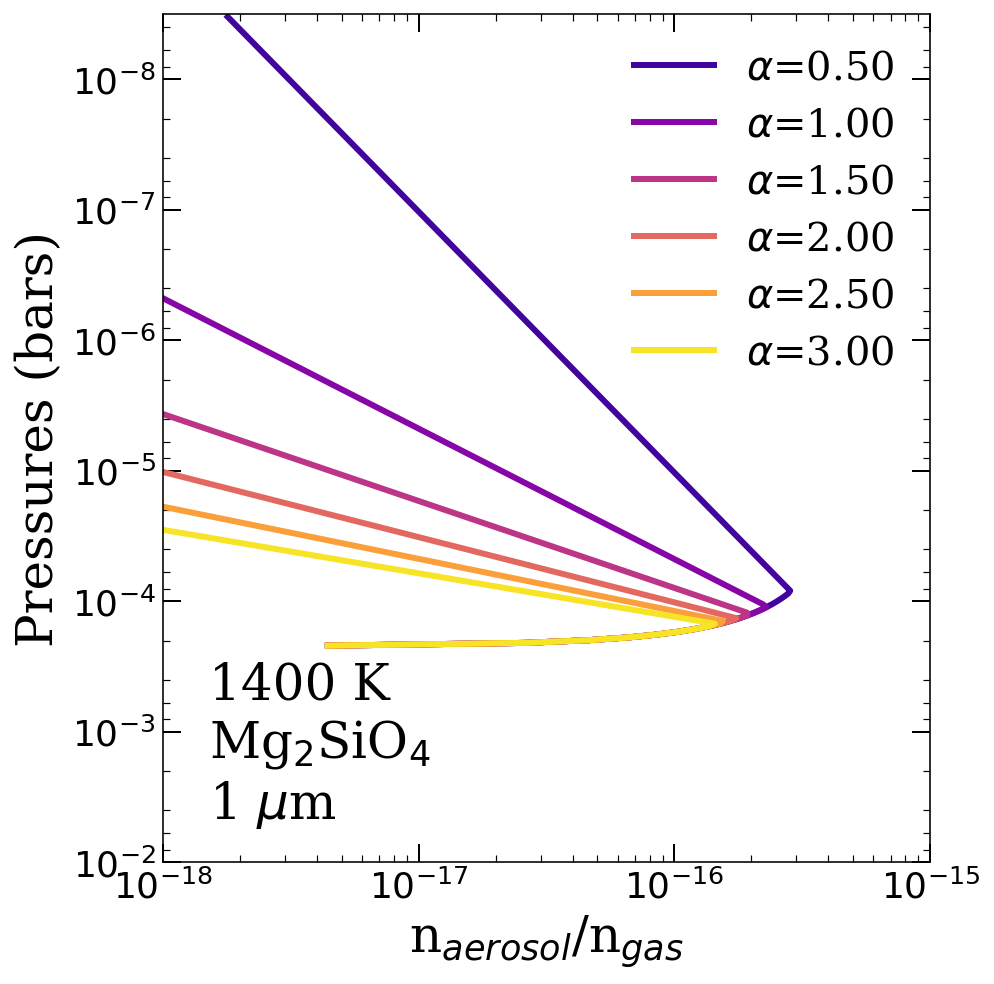}
    \includegraphics[width=0.475\textwidth]{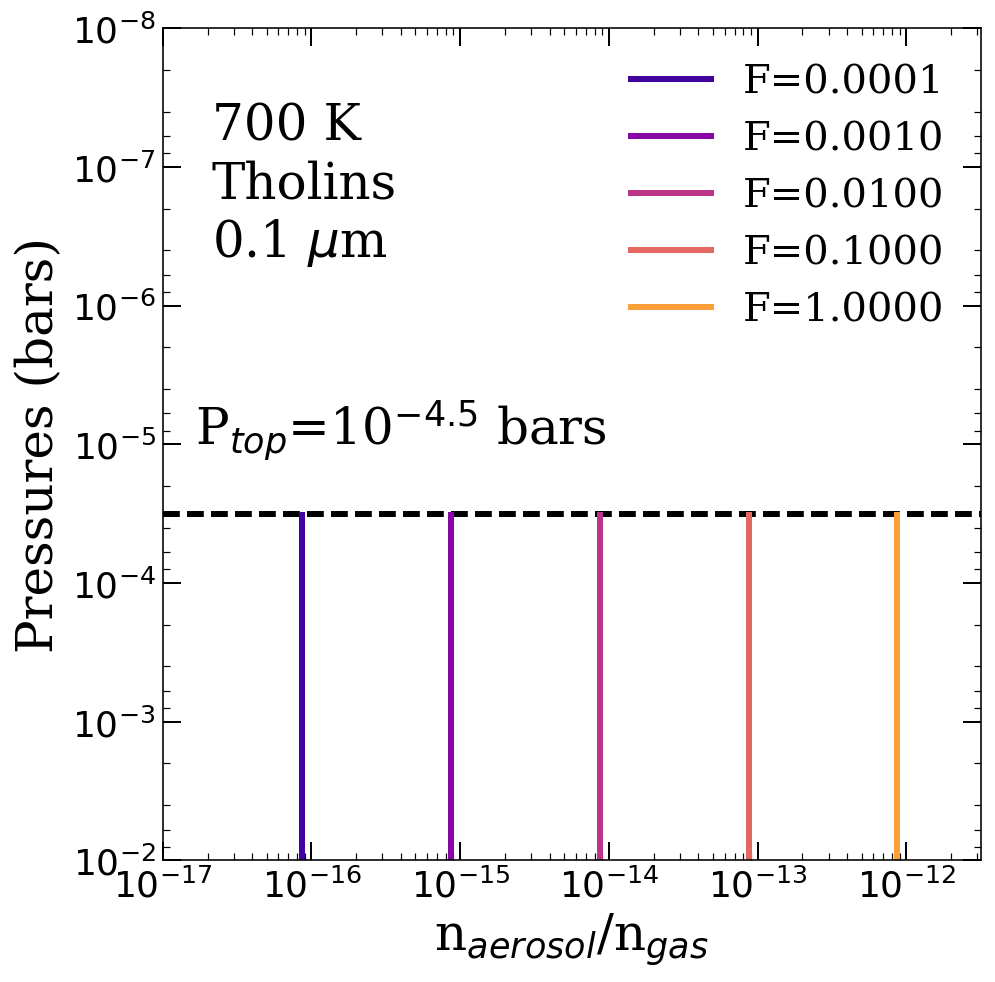}
    \caption{Demonstration of the meaning of parameters describing the spatial distributions of aerosols. On the left we show the equilibrium cloud with varying values of $\alpha$ and on the right we show the slab with varying values of $F$. The equilibrium cloud examples assume a forsterite cloud with 1 $\mu$m particles in a 1400-K atmosphere. The slab examples assume a Tholin haze with 0.1 $\mu$m particles in a 700-K atmosphere and a top-pressure cut-off of 10$^{-4.5}$ bars. In both cases we assumed the atmosphere had solar metallicity. }
    \label{fig:aerosol_spatial_dists}
\end{figure}

For our purposes, we want to use a small number of physically meaningful parameters that span a wide range of feasible aerosol behavior. For maximum flexibility, we separate these parameters into those that describe a particle size distribution and those that describe what we deem the ``aerosol spatial distribution." We adopt log-normal particle-size distributions throughout this work because they are simple, commonly used in atmospheric sciences, and can span a wide range of aerosol behavior. We place an upper limit on the amount of aerosol at a given layer in the atmosphere based upon the stoichiometry of an individual aerosol molecule, the atmosphere's metallicity, solar abundances of the least common constituent atom, and the density of that atmosphere layer. Note that a single aerosol particle will contain many, many aerosol molecules. We have already assumed homogeneous spheres in adopting Mie theory, so the relation between the number of molecules $N_{mol}$ and particle radius $a$ is just:
\begin{equation}\label{eq:nmol_a}
    N_{mol}(a) = (4/3 \pi a^3 \rho_{bulk}) / \mu_{mol} ~,
\end{equation}
where $\rho_{bulk}$ is the bulk density of the aerosol and $\mu_{mol}$ is the mass of a single molecule of the aerosol. 

It is not realistic that every single relevant atom will end up condensed in a cloud or photochemically incorporated into a haze. Therefore, we introduce some additional free parameters and assumptions. We focus on two flavors of aerosol spatial distribution: (1) a phase equilibrium option appropriate for condensing clouds, and (2) a slab aerosol which is appropriate for either a photochemical haze or a condensate species. The parameters for these two spatial parameterizations are summarized in Table \ref{tab:aerosol_spatial_parameters}, and explained visually in Figure \ref{fig:aerosol_spatial_dists}. 

For the slab aerosol, we simply include the aerosol from the bottom of the atmosphere all the way up to the top pressure cut off, which is specified by a free parameter P$_{top}$. We use an additional free parameter $F$ to dictate what fraction of the available ingredients are converted into aerosol particles, where $F$=1 implies every single one of the least abundant atoms are made into aerosol molecules. Note that this fraction is by number rather than mass. The interplay between $F$ and P$_{top}$ depends on the solar mixing ratio of the limiting atomic species, the metallicity of the atmosphere, and the particle size distribution. If $F$ is small, metallicity is small, and/or particle sizes are large, there will be a pressure below which the atmosphere has insufficient aerosol material to affect the transit spectrum. Changing P$_{top}$ to be lower than this pressure (i.e. higher in altitude) will have no effect on the transit spectrum. Conversely, if $F$, metallicity, and/or particle sizes are such that there is a large (i.e. optically thick) amount of aerosol material present at P$_{top}$, then changing $F$ may make little or no difference to the transit spectrum. Thus, for an optically thick slab aerosol, it is better to use just P$_{top}$, not $F$, but for an optically thin slab aerosols, both parameters may be needed to replicate the full range of aerosol behavior.

For the phase equilibrium cloud, we adopt the assumption that cloud bases occur where the Temperature-Pressure profile of a given atmospheric column roughly intersects the condensation curve for whichever species is being modeled. The condensate mixing ratio at the cloud base is then set by identifying the limiting gas involved in forming the chosen condensate and using up only that material which is in excess of the saturation vapor pressure. Moving up from the base, we take the minimum value between the material in excess of the saturation vapor pressure and the material that would be incorporated if we assume a smaller aerosol scale height than the gaseous pressure scale height given by:  

\begin{equation}\label{eq:tapering_eqCloud}
    \frac{n_{aerosol}}{n_{gas}} (P) = (\frac{n_{aerosol}}{n_{gas}})_{base}(P/P_{base})^{-\alpha}~.
\end{equation}

So a value of $\alpha$=0 keeps the ratio constant and a value of $\alpha$ greater than zero makes the number density of condensate particles fall off relative to the number density of gas particles. Studies of clouds in the Solar system gas giants have shown that the scale heights of clouds are typically three times smaller that gaseous pressure scale heights.  

\subsection{Computing which Pressures and Longitudes are Probed by Transit Spectra}\label{sec:contributions}

It can be useful to determine which pressure levels and longitudes impact the transit spectrum as the structure of the atmosphere and the properties of the aerosols in it vary. Our code lends itself well to determining which portions of the atmosphere are probed by transit spectra both by altitude and by longitude. If one computes the transit spectrum when the opacity in a single pressure level or in a single longitude slice is set to zero and finds a difference compared to the transit spectra of the full atmosphere, this indicates that the portion of the atmosphere in question is probed by the transit spectrum.

Using this method, we can compute the width of the transit limb by determining the maximum and minimum longitudes slices which cause a non-zero change in the transit spectrum. This can be split into a day-side width and a night-side width, since, as we will show in the following section, they are often not the same.

\section{Clear Transit Spectra of Hot Jupiters with Estimates for Day-side and Night-side Temperature} \label{sec:clear_paramstudy}

In our study, we use a parameterized structure to represent day-night variations (see \S \ref{sec:tgrad_structure}) introduced by \citealt{Caldas2019}. Our goal in this section is to provide the reader with an intuitive understanding of the transit spectra produced by atmospheres with this day-night temperature gradient parameterization when they are clear. This serves two purposes: (1) it lays the ground work for interpreting later results, and (2) it provides a sense of what types of objects will show significant effects of day-night temperature gradients in their transit spectra. The necessity of accounting for day-night temperature gradients when interpreting transit spectra varies as surface gravity, planet radius, and temperatures vary. It also depends on the $SNR$ and wavelength coverage of the data. We will use the sample of objects with phase curve observations presented in \citealt{Keating2019} to sample a realistic range of temperatures, surface gravities, and $SNR$. These objects have estimates for day-side temperature, night-side temperature, planet mass, planet radius, and stellar radius, so we only need to make assumptions for day-night transition width, reference pressure, and metallicity. Some of these objects have well-studied transit spectra, and some do not. The parameters used to model these objects are shown in Table \ref{tab:cowan_objects}. 

\begin{table}[]
    \centering
    \begin{tabular}{c|c|c|c|c|c|c}
        Name & T$_{Irr}$ &T$_{day}$&T$_{night}$& Mass& Radius & Stellar Radius\\
         & (K)&(K)&(K)&(M$_{J}$)&(R$_{J}$)&(R$_{\odot}$)\\
         \hline
HD189733b&1636&1279&979&1.142&1.138&0.805\\
WASP-43b&2051&1664&984&2.052&1.036&0.667\\
HD209458b&2053&1393&1015&0.690&1.380&1.203\\
CoRoT-2b&2175&1631&792&3.310&1.465&0.902\\
HD149026b&2411&1883&1098&0.357&0.718&1.497\\
WASP-14b&2654&2351&1267&7.341&1.281&1.306\\
WASP-19b&2995&2181&986&1.114&1.395&1.004\\
HAT-P-7b&3211&2678&1507&1.741&1.431&2.000\\
KELT-1b&3391&2922&1128&27.230&1.150&1.471\\
WASP-18b&3412&2894&815&10.430&1.165&1.230\\
WASP-103b&3530&2864&1528&1.490&1.528&1.436\\
WASP-12b&3636&2630&1256&1.470&1.900&1.657\\
WASP-33b&3874&3101&1776&2.100&1.603&1.444\\
    \end{tabular}
    \caption{Summary of properties for Hot Jupiters with measured day-side and night-side temperatures based on infrared phase curves \citep{Keating2019}.}
    \label{tab:cowan_objects}
\end{table}

\begin{figure}
    \centering
    \includegraphics[width=\textwidth]{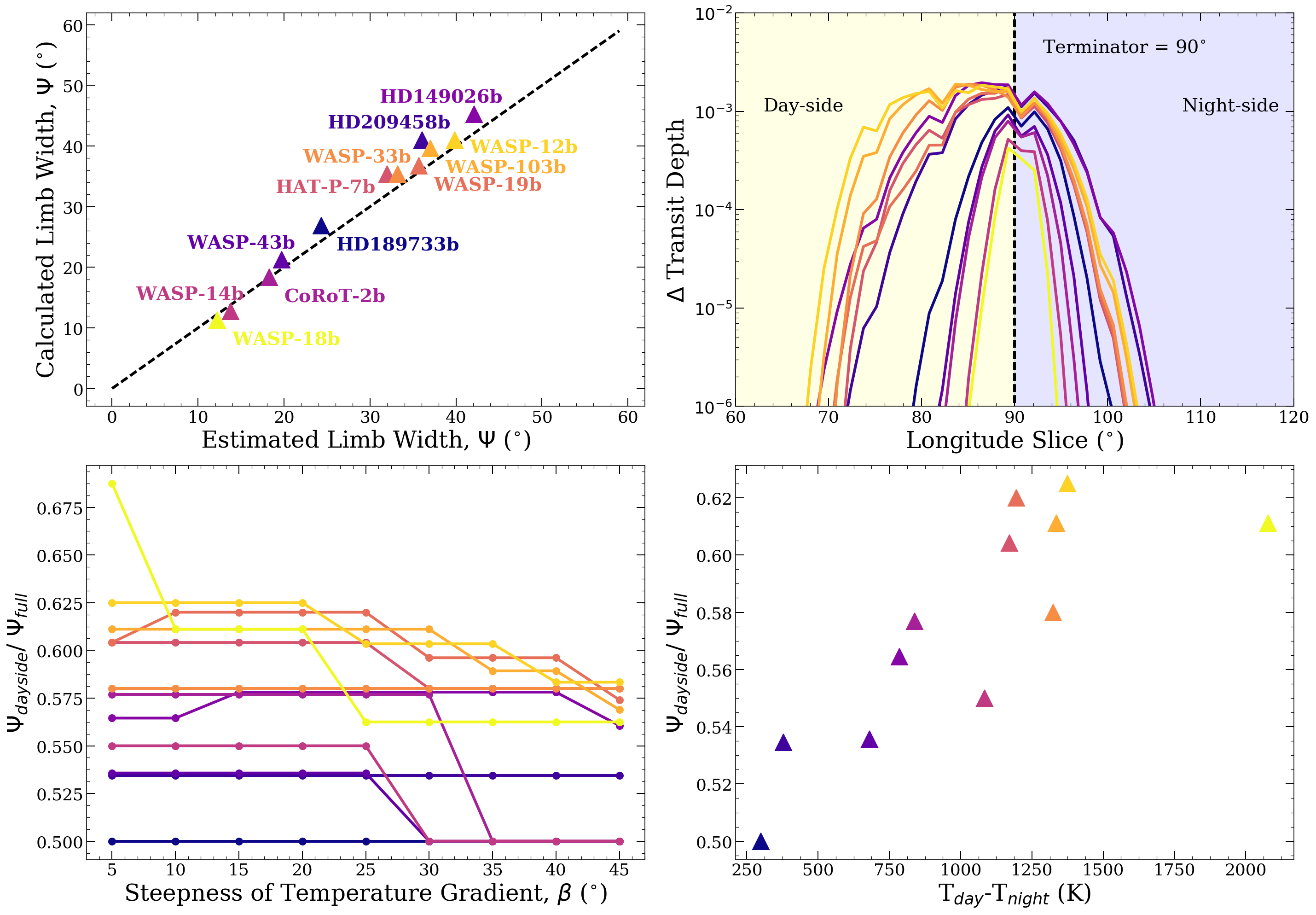}
    \caption{The top left panel compares our calculated limb widths to the order-of-magnitude estimate based on planet-radius and average scale height. The top right panel shows the profile of $\Delta$ Transit Depth versus longitude used to compute the limb widths. For the top row we assumed $\beta$=35$^{\circ}$. The bottom left panel shows how the width depends or does not depend on the steepness of the day-night transition. WASP-18b, which has a narrow limb width and the largest difference between day-side and night-side temperature, has the steepest shift towards being dominated by the day side as $\beta$ decreases. The bottom right panel shows how the fraction of the full width attributed to the day-side tends to scale with the difference between the day-side temperature and night-side temperature. We have assumed $\beta$=25$^{\circ}$ in this case. Note that in all panels the colors for each object are consistent, and that a sequential color scheme has been chosen and assigned based upon the magnitude of the difference between the day-side and nigh-side temperatures.}
    \label{fig:calculated_widths}
\end{figure}

\citealt{Caldas2019} demonstrated that for hotter, lower surface gravity objects (i.e. objects with large scale heights) the swath of atmosphere probed by transit spectroscopy ($\Psi$) can be surprisingly wide, using an order-of-magnitude calculation. As a first step towards assessing which objects might show the effects of day-night temperature gradients in their transit spectra, we perform this same order-of-magnitude estimate and a more detailed calculation of limb width (methodology outlined in \S \ref{sec:contributions}) for the objects in Table \ref{tab:cowan_objects}. The top left panel of Figure \ref{fig:calculated_widths} shows the excellent agreement between the estimates and more detailed calculations done using METIS. Nine out of the 12 objects have a limb width above 20$^{\circ}$, and the limb of HD149026b is as wide as 45$^{\circ}$. For tidally-locked and highly irradiated planets (such as those listed in Table \ref{tab:cowan_objects}), we expect the atmosphere to vary in temperature and chemical composition across the 10-45$^{\circ}$ swath of atmosphere probed by transit spectra \citep{Burrows2010}. The rest of Figure \ref{fig:calculated_widths} shows some of the interplay between the magnitude of the day-night temperature difference, the width of the day-night transition, limb width, and the degree to which the limb probed in transit skews towards the day-side of the planet.

 Because the day side of each planet will be hotter and have a larger scale height than the night side, the portion of the atmosphere probed in transit will be skewed towards the day-side (\citealt{Caldas2019}). The degree to which this occurs depends on the day and night temperatures, the width, $\beta$, over which the transition from day to night occurs, and the scale height of the planet. A smaller $\beta$ means the day-side takes up more of the total swath of atmosphere probed in transit (larger $\Psi _{day} / \Psi _{full}$). A larger day-night difference also tends to mean that the day-side takes up more of the total width.  The top right panel of Figure \ref{fig:calculated_widths} shows the change in average transit depth as the opacity in each longitude slice is set to zero. The dashed line marks 90$^{\circ}$, which is the day-night terminator. Looking at this, it is apparent that in many cases the contributions on the day side extend further away from the terminator than the night side. This panel assumes that $\beta$, which dictates the steepness/width of the day-night transition, is set to 35$^{\circ}$. The bottom left panel shows how this shift towards the day-side varies as $\beta$ changes from 5$^{\circ}$ to 45$^{\circ}$. The bottom right panel shows the ratio of day-side width to full width as a function of the difference in temperature between the day side and the night side when $\beta$=10$^{\circ}$. One can see that planets with a small difference between the day-side and night-side temperatures have a flatter relation between $\beta$ and the degree to which the day-side dominates the transit spectra.   

\begin{figure}
    \centering
    \includegraphics[width=\textwidth]{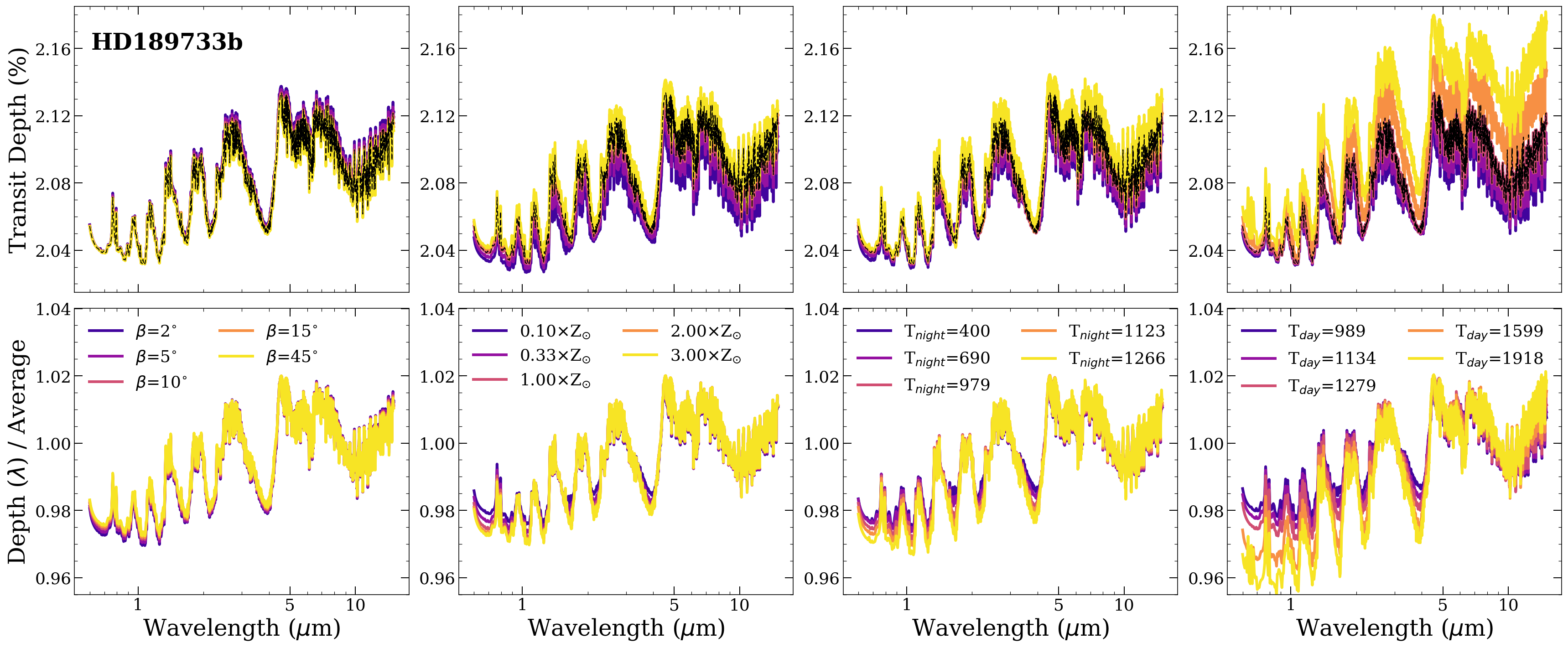}
    \caption{Demonstration of how HD189733b, one of the the six planets selected for MCMC retrieval experiments, is sensitive to $\beta$, $Z$, T$_{night}$, and T$_{day}$. The top row shows the resulting transit spectra as each parameter is perturbed about the fiducial values. The bottom row shows the same transit spectra divided by their average depths in order to highlight the change in shape rather than any shift up or down that is uniform across wavelengths.}
    \label{fig:HD189733b_clear_pstudy}
\end{figure}

\begin{figure}
    \centering
    \includegraphics[width=\textwidth]{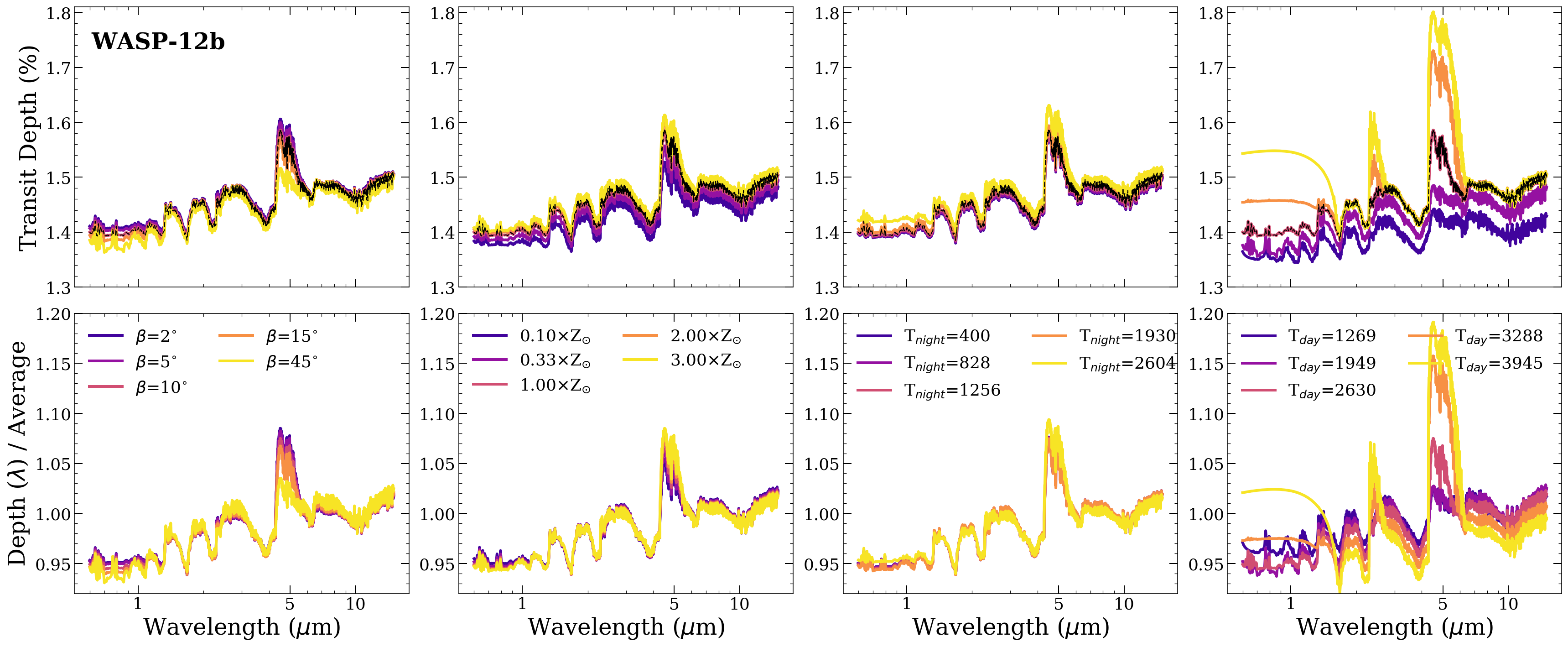}
    \caption{Demonstration of how WASP12b, one of the six planets selected for MCMC retrieval experiments, is sensitive to $\beta$, $Z$, T$_{night}$, and T$_{day}$. The top row shows the resulting transit spectra as each parameter is perturbed about the fiducial values. The bottom row shows the transit spectra divided by the average depth in order to isolate the change in shape rather than any shift up or down that is uniform across wavelengths.}
    \label{fig:WASP12b_clear_pstudy}
\end{figure}

Before we add aerosols and look at the results of the retrieval experiments, it will be informative to consider how the model parameters $\beta$, T$_{night}$, and T$_{day}$ affect transit spectra in comparison to $Z$. Figures \ref{fig:HD189733b_clear_pstudy} and  \ref{fig:WASP12b_clear_pstudy} show transit spectra for two of the objects from Table \ref{tab:cowan_objects} as the parameters are perturbed about fiducial values. HD189733b (Figure \ref{fig:HD189733b_clear_pstudy}) is the object with the smallest difference between the day-side and night-side temperature, while WASP-12b (Figure \ref{fig:WASP12b_clear_pstudy}) has one of the largest differences. In both figures the top rows show transit spectra as parameters are perturbed, while the bottom row shows the same transit spectra divided by their average depth to provide a sense of how the perturbations impact the \textit{shape} of the transit spectra isolated from any net shift up or down in depth. 

In the first column of Figures  \ref{fig:HD189733b_clear_pstudy} and  \ref{fig:WASP12b_clear_pstudy}, we perturb $\beta$. The degree to which perturbing $\beta$ impacts the transit spectrum provides one indicator of how important it might be to account for day-night temperature gradients when interpreting observations for a given object.  When the atmospheres are clear, there is some dependence on $\beta$ for both HD189733b and WASP-12b. As $\beta$ narrows, the spectra more closely resemble the spectra that would correspond to the day-side  temperature and as $\beta$ widens the atmosphere looks more like the spectra that would correspond to the mean of the day- and night-side temperature. A look at Figure \ref{fig:toymodel} and the bottom left panel of Figure \ref{fig:calculated_widths} makes the explanation for this clear. Broadening the transition region (raising $\beta$) means that the puffier day-side structure moves further from the limb of the planet, so less of it falls within the swath of atmosphere contributing to the transit spectra. In the limit of extremely large $\beta$, you see a transit spectrum that looks just like a uniform atmosphere with the mean temperature between day side and night-side. In the limit of extremely small $\beta$, you see simply the day-side temperature spectrum. This means that varying $\beta$ has a larger effect if there is a larger gradient between the day-side and the night-side (again we saw a foreshadowing of this in the bottom row of Figure \ref{fig:calculated_widths}). For the cooler planet, HD189733b, only the longer wavelengths vary slightly as $\beta$ changes. In these wavelength ranges the transit spectrum is formed higher up in the atmosphere than in the visible and NIR, so changing $\beta$ corresponds to a larger change in transverse path length. For the hotter planet, WASP-12b, varying $\beta$ has a large effect around the CO features at 4.5-5.5 $\mu$m and the metal hydrides in the optical and NIR. This reflects how changing $\beta$ shifts the spectra to being dominated by the higher temeprature day-side where equilibrium chemistry sets higher abundances of CO and metal hydrides. 

In the second column of Figures \ref{fig:HD189733b_clear_pstudy} and  \ref{fig:WASP12b_clear_pstudy}, we vary $Z$. For the cooler HD189733b, the relative strength of water and CO features compared to Rayleigh scattering and other opacity sources in the troughs and on the edges of water features will provide the constraint on metallicity. For the hotter WASP-12b constraints on metallicity will come from the relative strength of metal Hydrides in the optical and CO around 5 $\mu$m. One can immediately see that as $\beta$ varies between 5$^{\circ}$ and 45$^{\circ}$, it impacts the transit spectra on a similar level to perturbing the metallicity between $Z$ = 0.1$\times$ $Z_{\odot}$ and $Z$ = 3$\times$ $Z_{\odot}$ for WASP-12b. This indicates that ignoring a day-night temperature gradient would likely influence accuracy of the metallicity retrieved for such an object. However, for HD189733b, perturbing $Z$ has a steeper effect than $\beta$. Depending on the precision of the data, it may be that the presence of the day-night temperature gradient is not going to have a detectable effect on the transit spectrum. 

In the third column of Figures \ref{fig:HD189733b_clear_pstudy} and  \ref{fig:WASP12b_clear_pstudy}, we perturb T$_{night}$, and in the final column we perturb T$_{day}$. For HD189733b shifting T$_{night}$ has very similar effects to shifting the metallicity, although not identical. For WASP-12b, which has a transit spectra more heavily dominated by the day side of the planet, varying T$_{night}$ under 1256 K has very little effect. For both objects, varying T$_{day}$ has a larger effect than varying all the other parameters. For WASP-12b, it is especially striking when T$_{day}$ shifts to above 2000 K, which makes H$^{-}$ opacity important in the optical and causes CO absorption features strengthen significantly.  

In \S \ref{sec:mcmc_results}, we select half of the objects in Table \ref{tab:cowan_objects} for full MCMC experiments: WASP12-b, WASP103-b, HAT-P-7b, HD149026b, HD209458b, and HD189733b. These objects span the range of day-night temperature differences, and all have estimated limb widths above 25$^{\circ}$. We will see that the results of our retrievals reflect the trends elucidated in this section. When atmospheres are clear, the day-side temperature tends to have the tightest constraint, followed by the night-side temperature and then the metallicity. WASP-12b and WASP-103b, with their high temperatures and large differences between day-side and night-side, consistently have the tightest constraints on most parameters when the full temperature-gradient model is used for the retrieval. The cooler planets which exhibit the weakest dependence on $\beta$ (HD189733b and HD209458b) are not able to constrain $\beta$ at all, and so end up with a tail of hot day-side temperatures and cool night-side temperatures. This ends up biasing the retrieved metallicity to a value slightly lower than the true value. Retrievals of HAT-P-7b and HD149026b have the loosest constraints on all parameters, but this is due to their lower $SNR$'s, not their sensitivity to model parameters.

We will also see that the planets which exhibit a strong dependence on $\beta$ tend to have strong biases when the temperature gradient is ignored, but it can also be the case that goodness-of-fit indicators will make it clear that a single isothermal T-P profile is insufficient. The planets which exhibit the weakest dependence on $\beta$ have smaller biases (relative to their error bars) when we ignore the day-night temperature gradients, but goodness-of-fit indicators would leave us none the wiser.

\section{Combined Effects of Aerosols and Day-night Temperature Gradients} \label{sec:aerosol_paramstudies}
In this section we will examine how transit spectra change as we vary aerosol properties, metallicity, $\beta$, and day-side temperature. First we consider hazes, which extend from day-side to night-side, and then we look at condensed clouds, which only form at temperatures and pressures where our phase equilibrium cloud model allows. Our goal in this section is to evaluate whether we expect the presence of aerosols to mask or exacerbate the impact  of day-night temperature gradients on transit spectra and to anticipate any degeneracies between day-night gradient parameters, aerosol parameters, and metallicity which might impact in our retrievals. For all the transit spectra shown in this section, we assume a fiducial planet with a night-side temperature of 800 K, a mass of one Jupiter mass, a radius of 1.25 Jupiter radii at P$_0$ =  1 bar, $\beta$=15$^{\circ}$, and $Z$=1.05 $\times$ Z$_{\odot}$, orbiting a star the size of the Sun. The particle size distribution is a log-normal with $a_{m}$ = 1 $\mu$m and $\sigma _a$ = 2 when modeling clouds and a log-normal with $a_{m}$ = 0.1 $\mu$m and $\sigma _a$ = 2 when modeling hazes. 

\begin{figure}
    \centering
    \includegraphics[width=\textwidth]{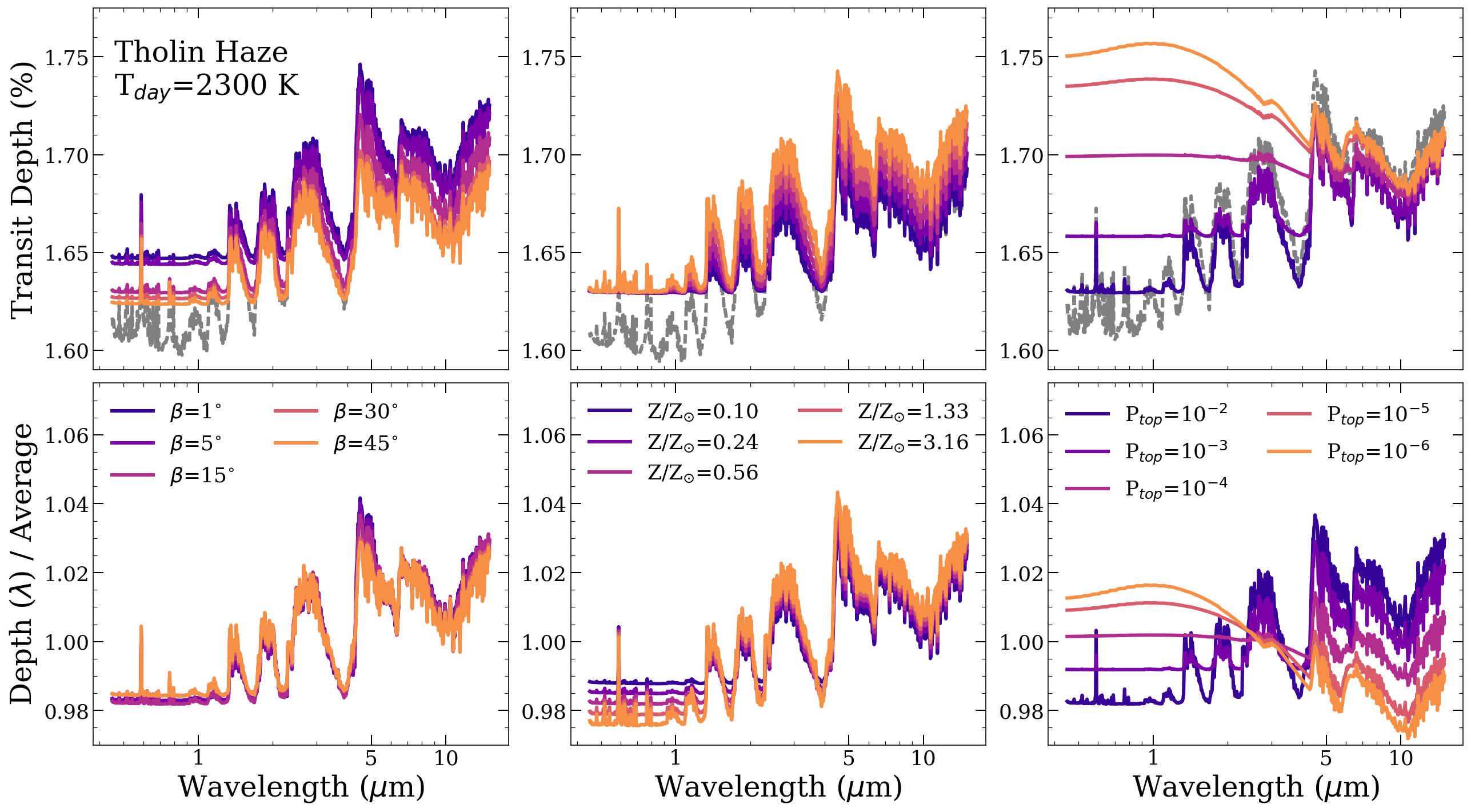}
    \caption{Transit spectra demonstrating the effects of perturbing $\beta$ (left column), $Z$ (center column), and $P_{top}$ (right column) about a fiducial case with a thick tholin haze. The top row shows transit spectra and the bottom row shows the same transit spectra divided by their average value to emphasize changes in shape rather than just shifts up or down across all wavelengths. We always assume a Jupiter mass planet with a radius of 1.25 Jupiter radii at P$_0$ = 1 bar, orbiting a solar radius star. The fiducial atmosphere and haze properties are: T$_{day}$ = 2300 K, T$_{night}$ = 800 K, $\beta$ = 15$^{\circ}$,  $Z$=1.05$Z_{\odot}$ (solar C/O ratio), P$_{top}$=10$^{-4}$ bars, $F$=0.25, and a log-normal particle size distribution with $a_{m}$ = 1 $\mu$m and $\sigma _a$ = 2.}
    \label{fig:hazy_pstudy}
\end{figure}

In Figure \ref{fig:hazy_pstudy}, we show hazy transit spectra with perturbations of $\beta$ (left column), $Z$ (center column), and $P_{top}$ (right column). The top row shows transit spectra and the bottom row shows the same transit spectra divided by their average value. In these spectra, the day-side temperature is assumed to be 2300 K and the haze is assumed to have the same complex indices of refraction as Titan tholins \citep{Khare1984}. For the fiducial combination of $Z$, particle size distribution, P$_{top}$=10$^{-2}$ bars, and $F$=0.25 chosen here, the tholin haze behaves like a gray opacity source because it is optically thick to all wavelengths at the pressure cut-off. Shifting $\beta$ affects the shape of the transit spectra a little, altering the strength of the CO feature relative to the haze at visible wavelengths (see bottom left panel). Perturbing $Z$ varies all the gaseous absorption strength relative to the haze which is stuck at the depth corresponding to the top-pressure cut-off. Moving the top pressure cut-off upwards to lower pressures relaxes the gray effect of the aerosol, dramatically changing the shape of the transit spectrum and encoding more information about the haze's particle size-distribution. When the haze is permitted to extend high up in the atmosphere (P$_{top}<$10$^{-4}$ bars), the spectral signatures of the tholins dominate over gaseous absorption at most wavelengths. From this figure, one can get a sense that, when the haze manifests as a gray opacity source, the information about $Z$ may be diluted as parameters become more degenerate. However, the bottom row of the figures shows that effect of varying $\beta$ is not totally degenerate with $Z$ and the top-pressure cut-off of the haze.

\begin{figure}
    \centering
    \includegraphics[width=\textwidth]{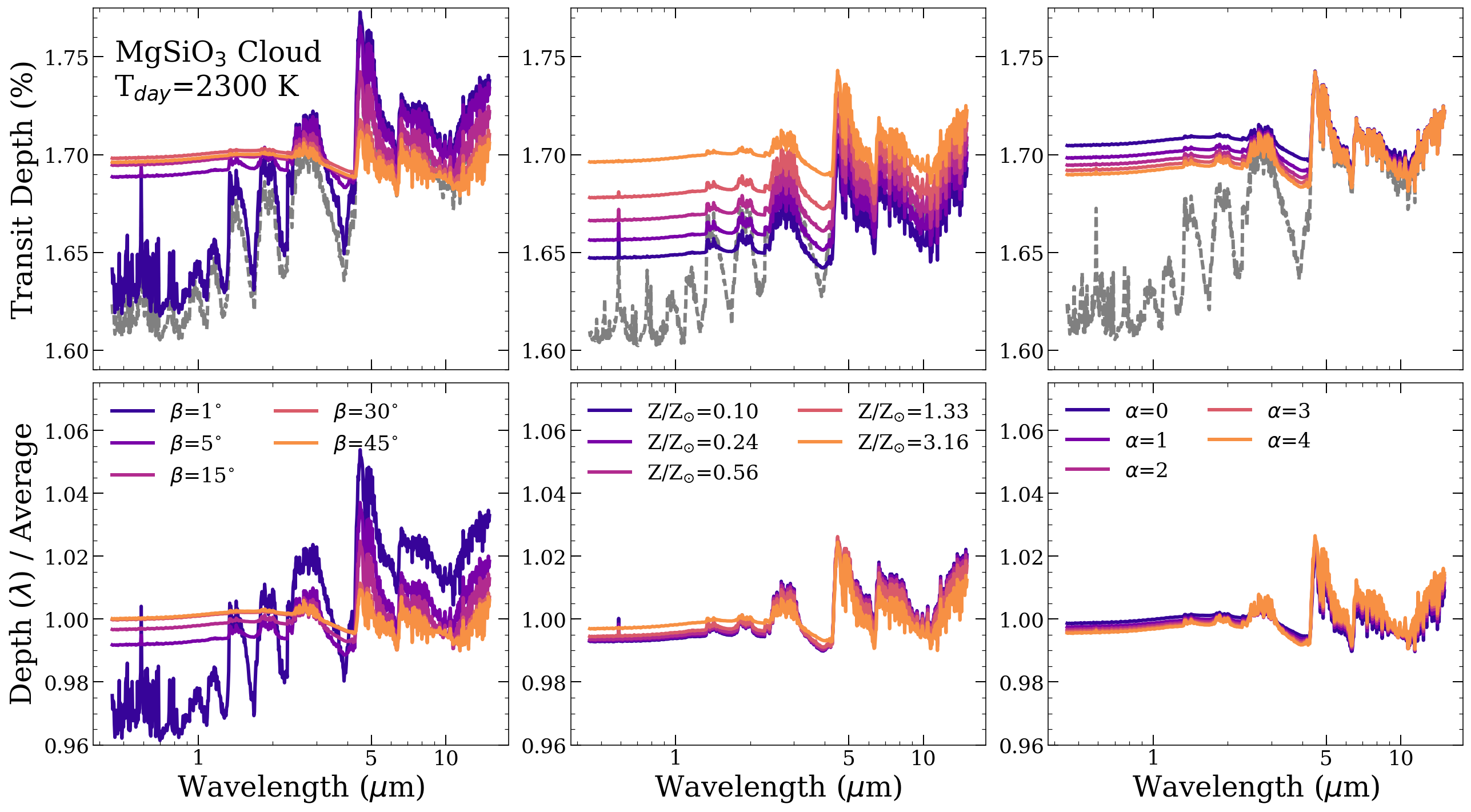}
    \caption{Transit spectra demonstrating the effects of perturbing $\beta$ (left column), $Z$ (center column), and $\alpha$ (right column) about a fiducial case with an enstatite cloud. The top row shows transit spectra and the bottom row shows the same transit spectra divided by their average value to emphasize changes in shape rather than just shifts up or down across all wavelengths. We always assume a Jupiter mass planet with a radius of 1.25 Jupiter radii at P$_0$ = 1 bar, orbiting a solar radius star. The fiducial atmosphere and haze properties are: T$_{day}$ = 2300 K, T$_{night}$ = 800 K, $\beta$ = 15$^{\circ}$,  $Z$=1.05$Z_{\odot}$ (solar C/O ratio), $\alpha$=2, and a log-normal particle size distribution with $a_{m}$ = 1 $\mu$m and $\sigma _a$ = 2.}
    \label{fig:cloudy_param_study}
\end{figure}

In Figure \ref{fig:cloudy_param_study}, we compare the effects of perturbing $\beta$, $Z$, and the ratio of the scale height of an equilibrium cloud relative to the gas pressure scale height ($\alpha$). This time, we perturb about a fiducial planet which still has T$_{day}$ = 2300-K day-side, but now it has an equilibrium cloud of enstatite with $\alpha$=2 instead of a tholin haze. Again, the top row shows transit spectra as the \% depth, and the bottom row shows the same spectra divided by their average value. Looking at the left column one can see that, under the assumptions of equilibrium cloud formation, varying $\beta$ can have dramatic effects on the transit spectra. Changing $\beta$ shifts a cloud's altitude and longitudinal position within the limb, and shapes the degree to which the limb is dominated by the day side. With clouds present, there is still a similar or even stronger dependence on $Z$ compared to a clear atmosphere. However we will see later that this can be degenerate with some other parameters in some cases. Varying $\alpha$ affects the transit spectra on a similar level to varying $Z$ in this particular case. Depending on the pressure where the cloud forms (i.e. the temperature and the species of cloud) and the particle size distribution, the degree to which varying $\alpha$ changes the transit spectrum can go from negligible to extremely significant.

\begin{figure}
    \centering
    \includegraphics[width=0.7\textwidth]{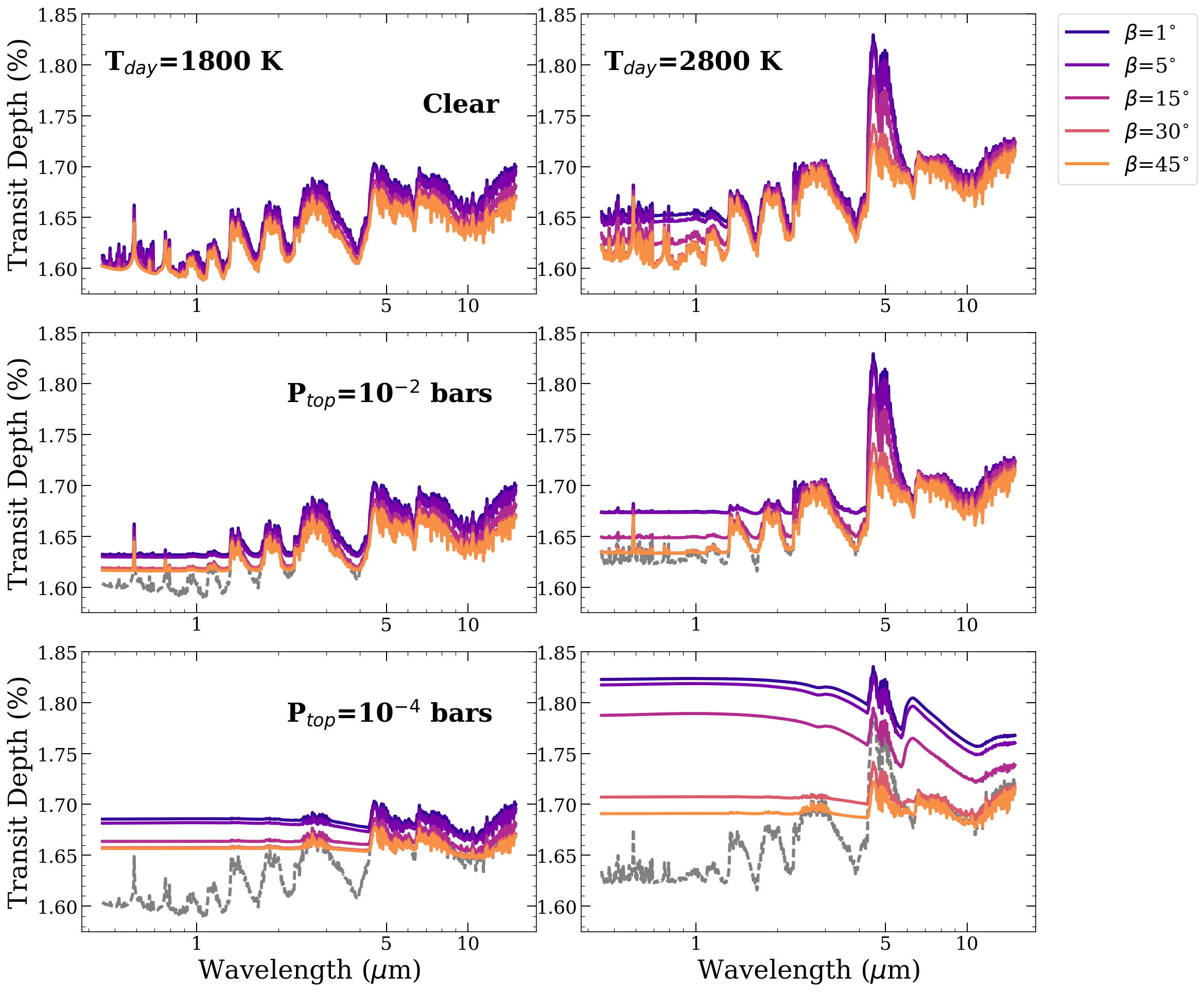}
    \caption{Demonstration of how transit spectra are sensitive to the width of the transition from day-side to night-side temperature ($\beta$). In the top row atmospheres are clear, in the middle row they include a tholin haze with P$_{top}$ = 0.01 bars, and in the bottom row they include a tholin haze with P$_{top}$=10$^{-4}$ bars. All atmospheres have a night-side temperature of 800 K, then each column shows results for a different day-side temperature increasing from left to right: 1300 K, 1800 K, 2300 K, and 2800 K. In the rows with aerosols included, we include the $\beta$=15$^{\circ}$ clear transit spectrum as a light gray dashed line.}
    \label{fig:3D_tholins}
\end{figure}

Figure \ref{fig:3D_tholins} shows how planets with hazes tend to exhibit greater sensitivity to the width of the day-night temperature transition region than clear atmospheres. The left column shows transit spectra for an atmosphere with a smaller day-side temperature of 1800 K and the right column shows transit spectra with a larger day-side temperature of 2800 K. Each row shows transit spectra with differing aerosol properties. The top row is totally clear, the middle row has a haze which only extends up to 0.01 bars, and the bottom row has a haze which extends up to 10$^{-4}$ bars. We have used the slab aerosol model, so the hazes extend uniformly across the day-night transition with no dependence on temperature. The colors indicate the width of the region where the temperature changes from the day-side temperature to the night-side temperature. When aerosols are present, their additional opacity raises the altitude shaping the transit spectrum, so a change in $\beta$ now corresponds to a larger change in transverse path length, especially at optical wavelengths. For the planet with an 1800-K day-side temperature, when it is clear, there is very little dependence on $\beta$, but, when it has a haze at altitude, there is a much steeper dependence. The planet with a 2800-K day-side temperature already exhibits a fairly strong dependence on $\beta$ when clear. This dependence becomes even steeper for the high altitude haze case. It seems that incorporating a haze that spans the day-side and night-side will likely exacerbate rather than eliminate the need to account for day-night temperature gradients when interpreting transit spectra.

\begin{figure}
    \centering
    \includegraphics[width=0.7\textwidth]{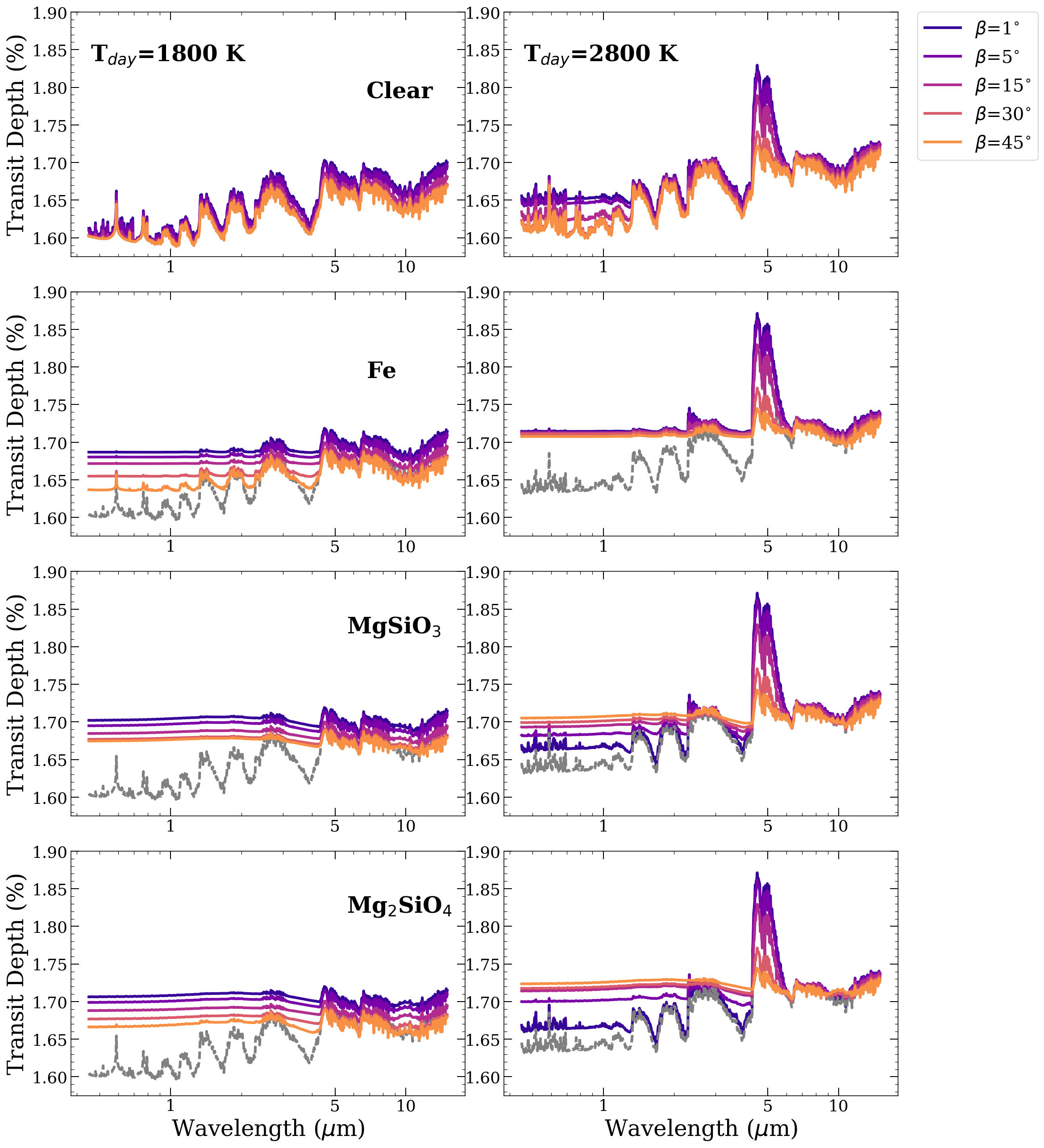}
    \caption{Demonstration of how varying the width of the day-night transition ($\beta$) changes cloudy transit spectra with a variety of cloud species (Fe in second row, MgSiO$_3$ in third row, Mg$_2$SiO$_4$ in bottom row), and with two different day-side temperatures (1800 K left column, 2800 K right column). The transit spectra always assume a Jupiter mass planet with a radius of 1.25 Jupiter radii at P$_0$ = 1 bar, orbiting a solar radius star. The atmospheres have T$_{night}$ = 800 K and  $Z$=1.05$Z_{\odot}$ (solar C/O ratio). The clouds have $\alpha$=2, and a log-normal particle size distribution with $a_{m}$ = 1 $\mu$m and $\sigma _a$ = 2. }
    \label{fig:cloudy_beta_study}
\end{figure}

In Figure \ref{fig:cloudy_beta_study}, we show transit spectra for the same 1800-K day side and 2800-K day side objects as in Figure \ref{fig:3D_tholins}. This time we compare the effects of perturbing $\beta$ for a clear atmosphere (row one) to atmospheres with a phase equilibrium cloud of Fe (row two), MgSiO$_3$ (row 3), and Mg$_2$SiO$_4$ (row 4). For each combination of species, day-side temperature and $\beta$, the equilibrium cloud has formed in a different range of pressures and longitudes. For the cooler day side of 1800-K, adding in clouds always increases the dependence on $\beta$ dramatically, especially from 0.5-5 $\mu$m. In this case, Fe condenses deeper in the atmosphere at $\beta$ narrows. When Fe condenses too high up in the atmosphere there is not much material, so the cloud is optically thin and still lets a lot of gaseous opacity contribute (orange line in left column second row). For the warmer day side of 2800-K, the Fe cloud actually decreases the dependence on $\beta$ as it masks the effects of the hydrides. In this case, the Fe cloud will form across a wider swath of the limb compared to the 1800-K planet. For the two types of silicate cloud (MgSiO$_3$ and Mg$_2$SiO$_4$), there is a very different dependence on $\beta$ between the 1800-K day-side planet and the 2800-K day-side planet. The silicates only condense further from the day-side of the hotter planet, so, if $\beta$ gets too small, the cloud shifts almost entirely out of the limb and its effect on the transit spectra diminishes. Whereas, for the 1800-K day-side planet the cloud is always forming across the full extent of the limb. These parameter sensitivity-studies indicate that imposing equilibrium assumptions on clouds usually leads to a stronger sensitivity to T$_{day}$, T$_{night}$, and $\beta$ than for a clear atmosphere.

In our retrieval experiments for cloudy and hazy atmospheres with temperature gradients we will get results consistent with the implications of these parameter studies. When a haze is included, there is generally a stronger constraint on T$_{day}$, T$_{night}$, and $\beta$ than there would be for a clear atmosphere. However, the metallicity information is sometimes scrambled in with aerosol properties and lost, especially if the haze manifests as a gray opacity (note, this would happen irregardless of the presence of a day-night temperature gradient). When an equilibrium cloud is included, the same stronger-than-clear constraint on T$_{day}$, T$_{night}$, and $\beta$ is found, and, at times, a constraint can still be placed on $Z$. The equilibrium cloud form of aerosol can still manifest as a gray opacity source, but it happens much less frequently than for the slab form of aerosol.

\section{Retrieval Experiments}\label{sec:mcmc_results}

In this section we present results for MCMC retrieval experiments testing whether there is sufficient information in transit spectra to constrain a  model which includes day-night transition properties, and how ignoring day-night temperature gradients may bias retrieval efforts. To explore a range of parameter space, we simulated data for half of the twelve hot Jupiters in Table \ref{tab:cowan_objects} for three different cases: when they are clear, when they have a phase equilibrium MgSiO$_3$ cloud, and when they have a slab aerosol of MgSiO$_3$. We then fit these data using a model with a single isothermal T-P profile and using a model with a day-night temperature gradient (see \S \ref{sec:mcmc_methods} for a more detailed description of retrieval procedures, and \S \ref{sec:noise_model} for a description of how we simulated approximate JWST-like observations). For comparison, we also simulated data for a uniform temperature atmosphere and fit this with a single isothermal T-P profile. The six objects are: HD189733b, HD209458b, HD149026b, HAT-P-7b, WASP-103b, and WASP-12b. These objects 1) span a variety of $SNR$'s (see Figure \ref{fig:noise}), 2) span the range of day-night temperature gradient magnitude, and 3) were chosen to have preferentially larger than 20$^{\circ}$ limb widths (see Figure \ref{fig:calculated_widths}). In the rest of this section we describe the results of these retrievals. We begin with clear atmospheres in \S \ref{sec:clear_fits}. Then we move on to atmospheres with a uniform haze in \S \ref{sec:hazy_fits} and a night-side condensed cloud in \S \ref{sec:cloudy_fits}. Similar experiments have been done before, but only for clear atmospheres with uniform chemistry between the day and night side \citep{Caldas2019} or clear atmospheres with just H$^{-}$ opacity varying \citep{Pluriel2020}. Our study adds to these previous works by varying the abundances of all the major opacity sources across the terminator in accordance with thermochemical equilibrium and considering cases where aerosols are present.

\subsection{Clear Atmospheres}\label{sec:clear_fits}
We begin with clear atmospheres to evaluate how previous results hold up when the abundances of all the significant opacity sources are assumed to vary, and to provide a point of comparison for the subsequent studies which incorporate clouds and hazes. This is also the first study to directly try to fit a model with a day-night temperature gradient. This allows us to evaluate whether spectra could contain sufficient information to actually constrain separate day- and night-side temperatures and constrain how quickly or slowly the atmosphere transitions from one to the other. Such information could inform our understanding of heat-redistribution and circulation on tidally-locked exoplanets, providing a useful link between phase curve observations and transit observations.

\begin{figure}
    \centering
    \includegraphics[width=\textwidth]{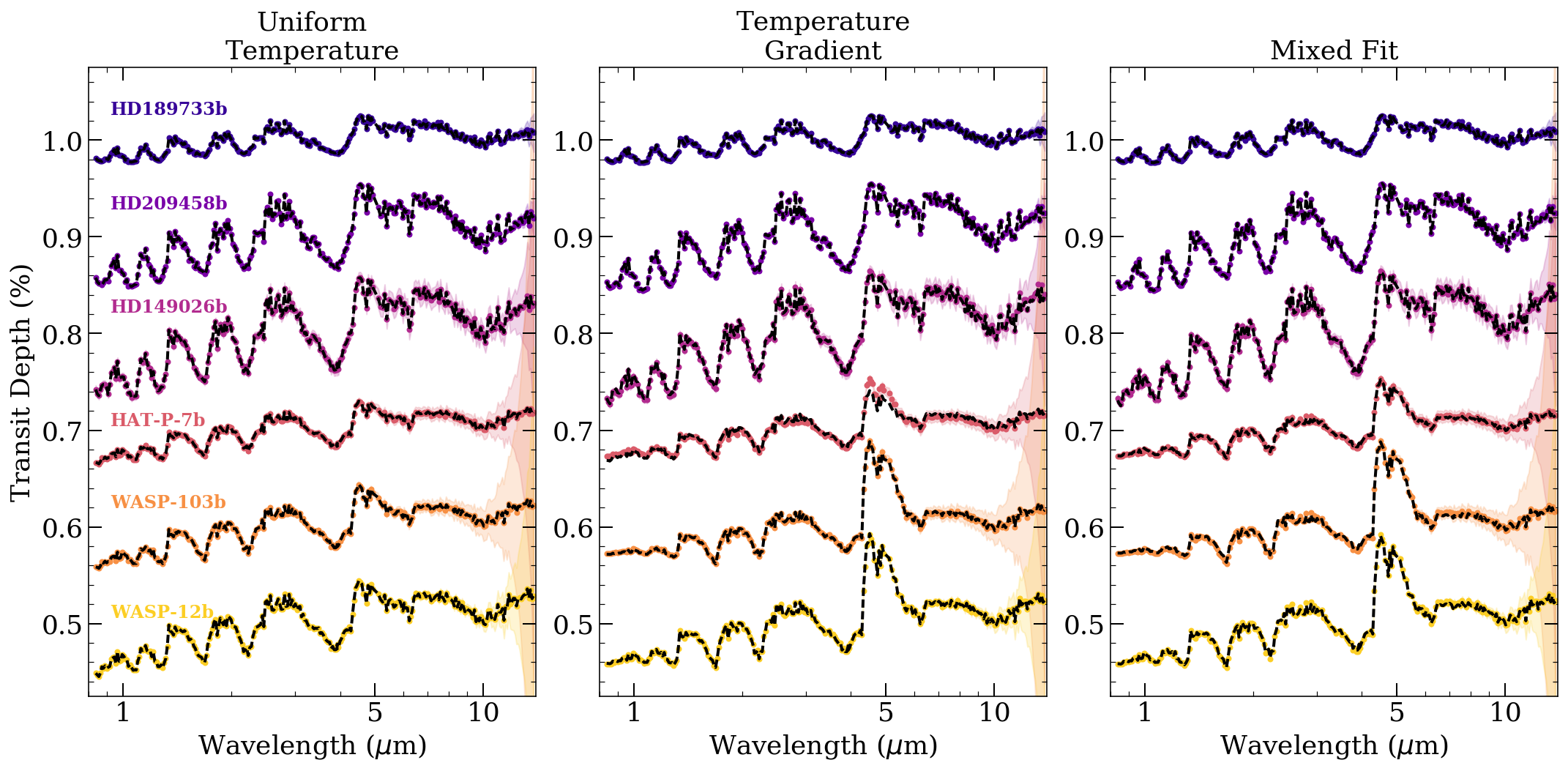}
    \caption{Transit spectra corresponding to the median values of posteriors (black dashed lines) compared with clear simulated data (colored dots with shaded error envelope) in our MCMC retrieval experiments. Note that fits and simulated data have been divided by their average value and off set from each other by a constant value. The data in the left panel was simulated for an atmosphere with a uniform temperature and fit using a uniform temperature model, the center column has a temperature gradient in the data and is fit accordingly, and the right column has a temperature gradient in the data but is fit with a uniform temperature model. All the atmospheres are clear with $Z$=1.05$\times Z_{\odot}$ and P$_{0}$=1 bar. When a temperature gradient is present, $\beta$=10$^{\circ}$. The temperatures, masses, planet radii and stellar radii are taken from Table \ref{tab:cowan_objects}. When a single temperature is used to generate data, it is the average between the day-side temperature and night-side temperature in Table \ref{tab:cowan_objects}. }
    \label{fig:clear_bestfits}
\end{figure}

Figure \ref{fig:clear_bestfits} shows the simulated transit spectra (colored dots with shaded error envelope) along with the model corresponding to the median values of the posterior distributions mapped out by the MCMC retrievals (black dashed lines). The left panel shows the control case: a uniform temperature atmosphere fit by a single T-P profile. The center panel shows the spectra for atmospheres with day-night temperature gradients fit by a model which includes a day-night temperature gradient. The right panel shows the test-case: data simulated for an atmosphere with a day-night temperature gradient, but fit using a single isothermal T-P profile. One can readily see that the best-fit spectra mostly fit the data very well. This agrees with the finding of \citealt{Caldas2019} that, for clear atmospheres, there is not an indication from typical goodness-of-fit estimators that a single T-P model is inadequate, even if it is causing biased results. The only apparent discrepancy is actually when HAT-P-7b is fit by a model which accounts for the day-night temperature gradient. The best-fit spectra underestimates the strength of the CO feature around 4.5-5.5 $\mu$m.

\begin{figure}
    \centering
    \includegraphics[width=\textwidth]{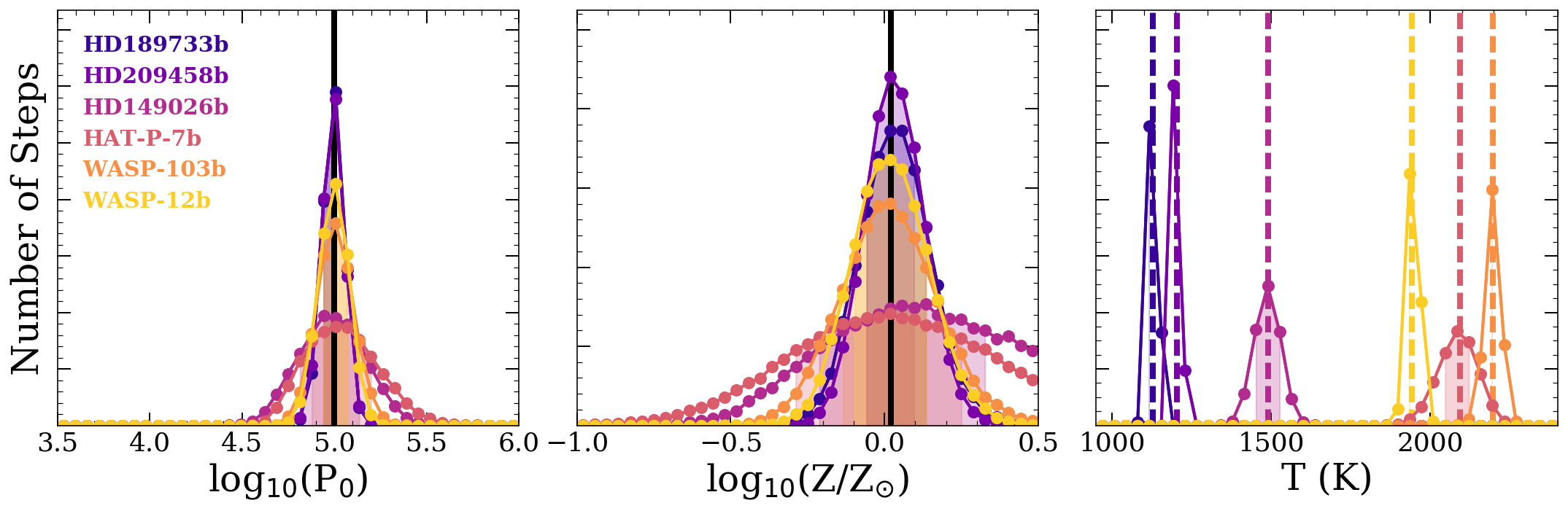}
    \caption{The posteriors for the clear, uniform-temperature control case. Each set of colored dots shows the histogram for a different object. The black lines mark the true values of the reference pressure, P$_0$, and the metallicity $Z$. Different objects have different temperatures, so the true temepratures are marked with colored dashed lines in the right panel. Light shading marks the region which falls between the sixteenth and eighty-fourth percentile of the posterior.}
    \label{fig:clear_hists_uniform}
\end{figure}

In the clear uniform temperature control case, all the retrieved values are very accurate and the posteriors are approximately Gaussian (see Figure \ref{fig:clear_hists_uniform}). The varying width of the posteriors track with the varying $SNR$ of the targets. These same variations in $SNR$ will affect the rest of the fits as well. 

\begin{figure}
    \centering
    \includegraphics[width=\textwidth]{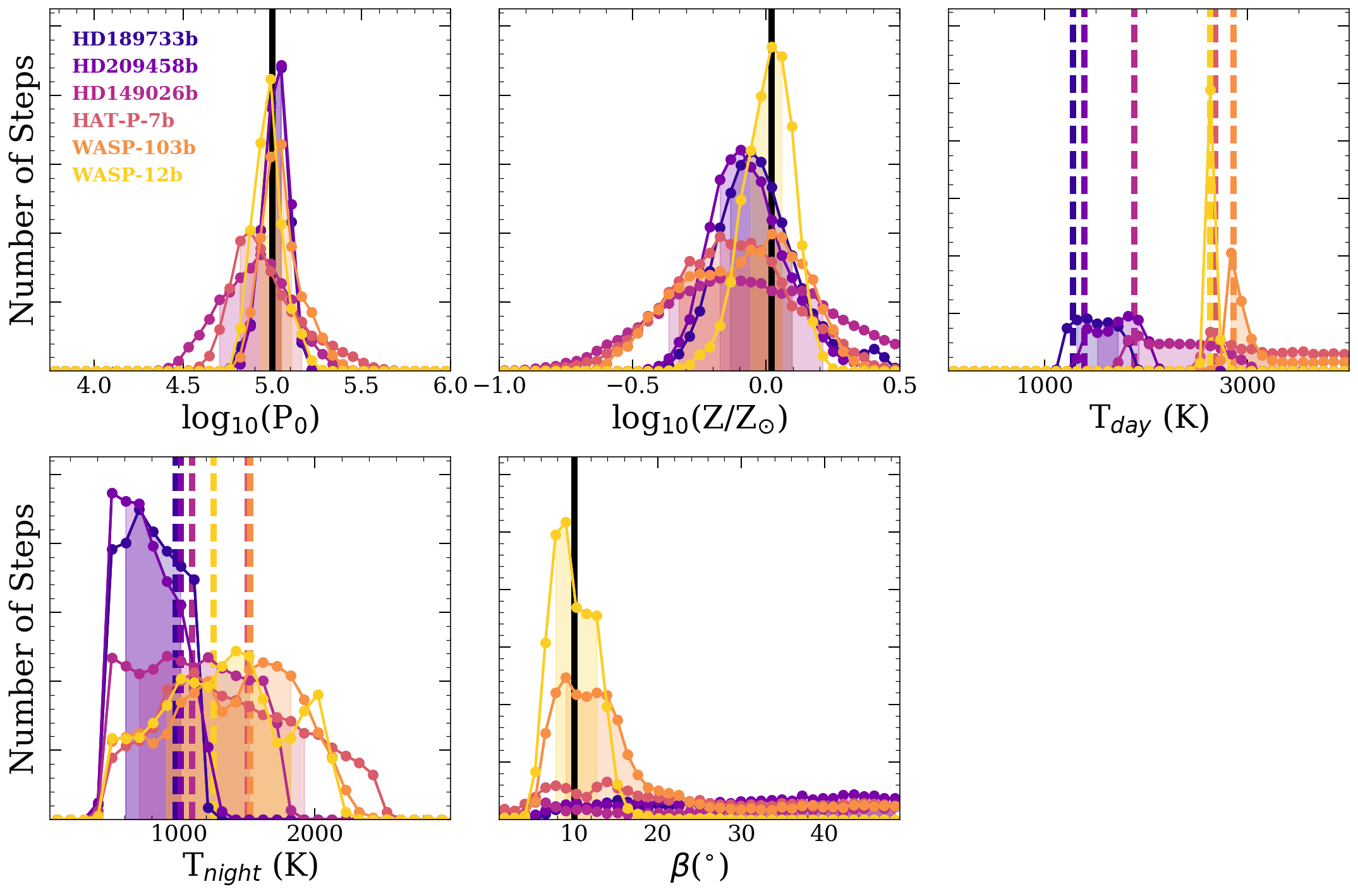}
    \caption{The posteriors for the clear control case with a day-night temperature gradient. Colored dots and lines show histograms of the MCMC chains for each object. Shaded regions mark the sixteenth through eighty-fourth percentile. Vertical lines indicate the true values of parameters. The meaning of the parameters are described in detail in \S \ref{sec:tgrad_structure}.}
    \label{fig:clear_hists_tgrad}
\end{figure}

In the clear case fit with a temperature gradient (Figure \ref{fig:clear_hists_tgrad}), we see that some objects fall in an area of parameter space with large degeneracies between $\beta$, T$_{day}$ and T$_{night}$ (HD189733b, HD209458b, HD149026b, and HAT-P-7b), while other spectra are able to constrain the more complex model (WASP-103b and WASP-12b). The correct day-side temperature is always within the 16-84 percentile range of the posterior, but there is a long tail towards higher temperatures. The night-side temperature posteriors always contain the true value within the 16-84 percentile range, but it also has a tail, this time tending towards lower temperatures. These skews towards high day-side temperatures and low night-side temperatures correspond to the very long tail in the posterior of $\beta$, extending up towards large angles. This degeneracy simply reflects the geometry of our toy-model and the finite angular extent of the atmosphere probed by transit spectroscopy. If you lower T$_{night}$ and raise T$_{day}$, but broaden $\beta$ you can get the same range of temperatures with longitude to fall within the region about the terminator probed by the transit spectrum. For WASP-12b and WASP-103b, there is an actual constraint placed on $\beta$, so it seems the CO feature which is formed at higher altitudes in hot atmospheres is key to cutting off the long tail of $\beta$ towards high values. The metallicities retrieved for the day-night temperature gradient control case are slightly biased towards lower metallicities in the objects which don't have a constraint on $\beta$, but still agree with the true metallicity within their error bars. WASP-12-b actually has a tighter metallicity constraint in the temperature gradient case than in the uniform temperature case. WASP-103b has a multi-modal metallicity posterior with one peak centered on the correct metallicity.  

\begin{figure}
    \centering
    \includegraphics[width=\textwidth]{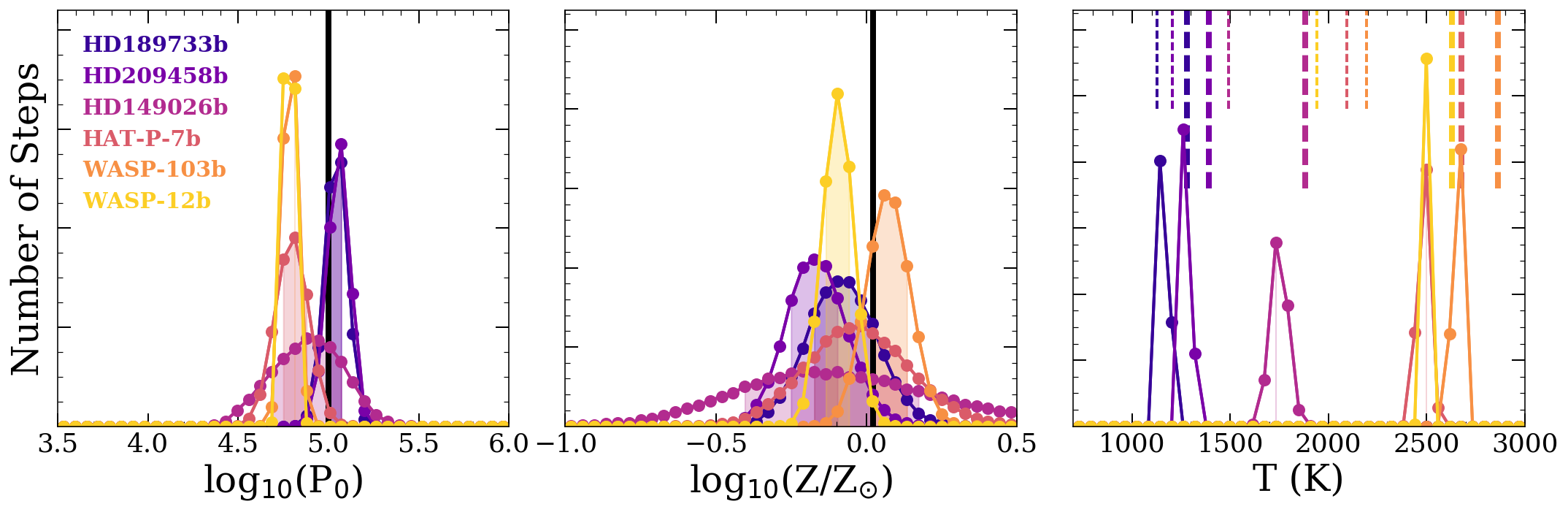}
    \caption{Posteriors for the clear test case where a uniform-temperature model is used to fit data that was simulated using a day-night temperature gradient. Colored dots and lines show histograms for each object. Black lines mark the true values of P$_0$ and $Z$. There is not a well-defined single true temperature for such data, so we have included thinner shorter lines marking the average between the day- and night-side temperatures and longer thicker lines marking the day-side temperatures for each planet.
    }
    \label{fig:clear_hists_mixed}
\end{figure}

Figure \ref{fig:clear_hists_mixed} shows all the posteriors for the clear test case, where we have fit a temperature gradient with a single T-P profile. The retrieved temperatures fall between the day- and night-side temperature, favoring the day-side very strongly for HD149026b, HAT-P-7b, WASP-12b and WASP-103b. This is consistent with results reported in \citealt{Caldas2019}, and with what we expect from our calculations demonstrating that the day-side contributes more to the transit spectrum than the night-side (Figure \ref{fig:calculated_widths}). For HD189733b and HD209458b the retrieved temperature is much closer to the average of the day-side and night-side temperatures. The retrieved values of $P_{0}$ and $Z$ deviate from the ground truth in an attempt to mimic the effects of the day-night temperature gradient, while forced to use a single temperature. Metallicity is important because it is tied to planet formation theories, but there are other parameters we may be misunderstanding. None of the metallicity posteriors peak at the true value. For HD209458b and WASP-12b the 16-84 percentile range excludes the true metallicity entirely. For HAT-P-7b and HD209458b the true value falls right at the edge of the 16-84 percentile range. For HD149026b and HAT-P-7b, with the lowest $SNR$, the 16-84 percentile range includes the true metallicity. For HD189733b, $P_{0}$ is biased slightly high and $Z$ is biased slightly low, but the values just barely agree with the true values within the error bars. This object has a relatively small day-night temperature gradient but a high $SNR$, so it is right at the boundary of having sufficient $SNR$ that accounting for the effects of its small day-night gradient matters. It is definitely not necessary to account for the day-night temperature gradient in HD149026b. Other objects retrieve the wrong value for one or both of $P_{0}$ and $Z$, so it will be necessary to account for day-night temperature gradients to get unbiased results. 

The clear MCMC experiment results indicated that the presence of a day-night temperature gradient on a tidally-locked hot Jupiter has a significant enough effect on its transit spectrum to bias our retrieval results when it is unaccounted for, provided the $SNR$ is not too low, and the difference in temperature between the day-side and night-side is larger than around 400 K or so. These findings are consistent with previous studies. We have also found that, for some planets, there is sufficient information encoded in the transit spectrum to constrain a separate day-side and night-side temperature.

\subsection{Hazy Atmospheres}\label{sec:hazy_fits}

Now we show results for the same MCMC experiment using transit spectra of planets with a thick enstatite slab aerosol in their atmosphere. This haze is present across the day- and night- side extending up to pressures of 10$^{-4.5}$ bars. The thick, high altitude haze introduces some new degeneracies between $Z$ and aerosol properties like $F$ and a$_m$. It also strengthens the degeneracy between $Z$ and P$_0$ since key gaseous features in the optical wavelength range are obscured. Adding in aerosols always complicates matters regardless of the presence of a day-night temperature gradient. This is well-established by previous retrieval studies. In this sub-section and the following sub-section \ref{sec:cloudy_fits}, we build up from there to explore interplay between aerosols and day-night temperature gradients. Are degeneracies exacerbated or broken? If day-night temperature gradients are not accounted for in hazy atmospheres, do we see similar biases in the retrieved metallicities as has been noted for clear atmospheres? 

\begin{figure}
    \centering
    \includegraphics[width=\textwidth]{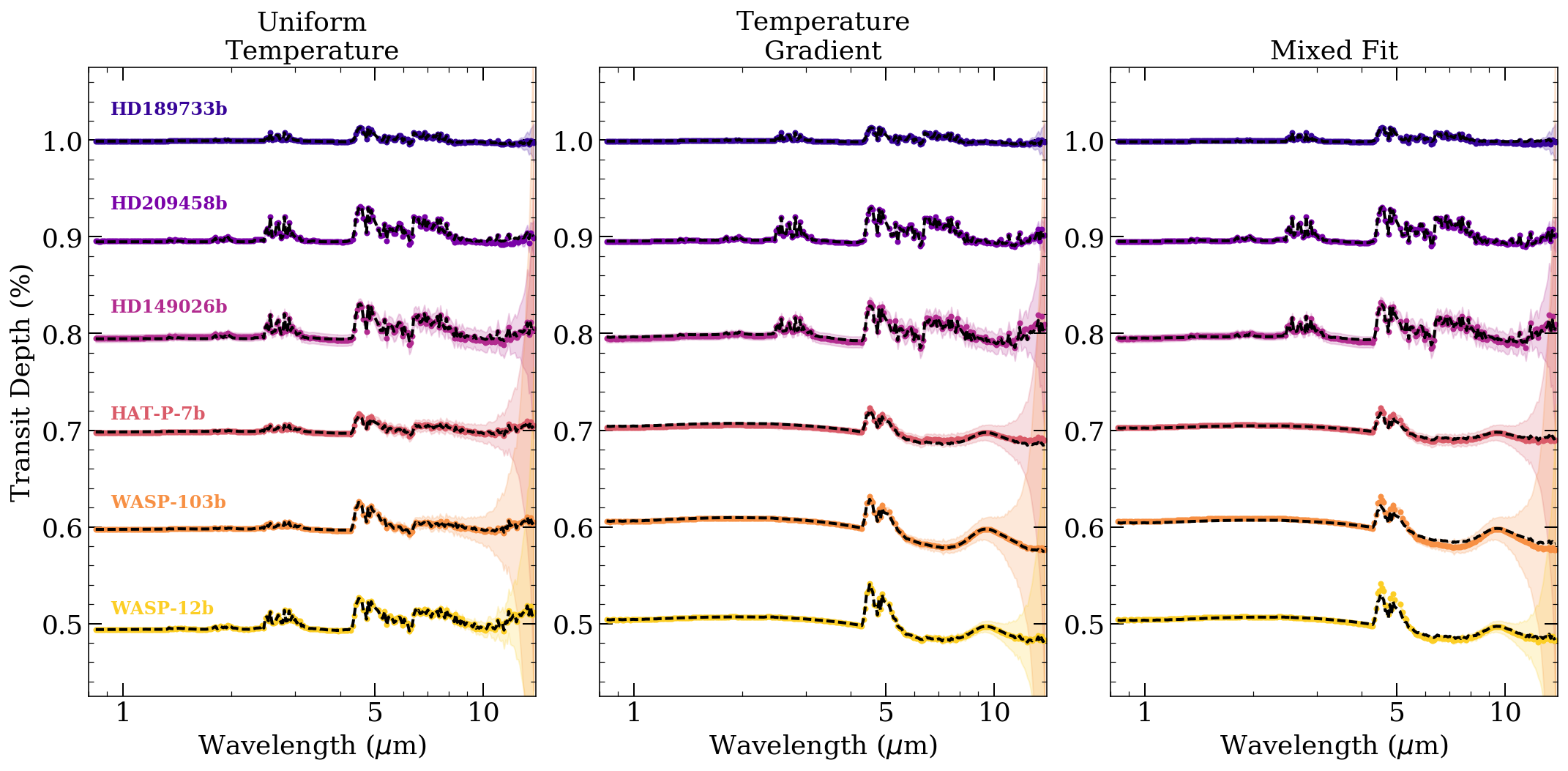}
    \caption{The hazy simulated data (colored dots with shaded error envelope) and best fit spectra (black dashed lines) taken from the median values of MCMC chains. The left panel shows the uniform-temperature control case, the center panels shows the control case with a day-night temperature gradient, and the right panel shows the test case where data was simulated using a temperature gradient but then fit with a uniform-temperature model. All models included a slab aerosol of MgSiO$_3$ with a modal size of 1 $\mu$m, a size dispersion of 1.75, a top-pressure cut-off of 10$^{-4.5}$ bars, and half of the available material incorporated into particles.}
    \label{fig:hazy_bestfits}
\end{figure}

\begin{figure}
    \centering
    \includegraphics[width=\textwidth]{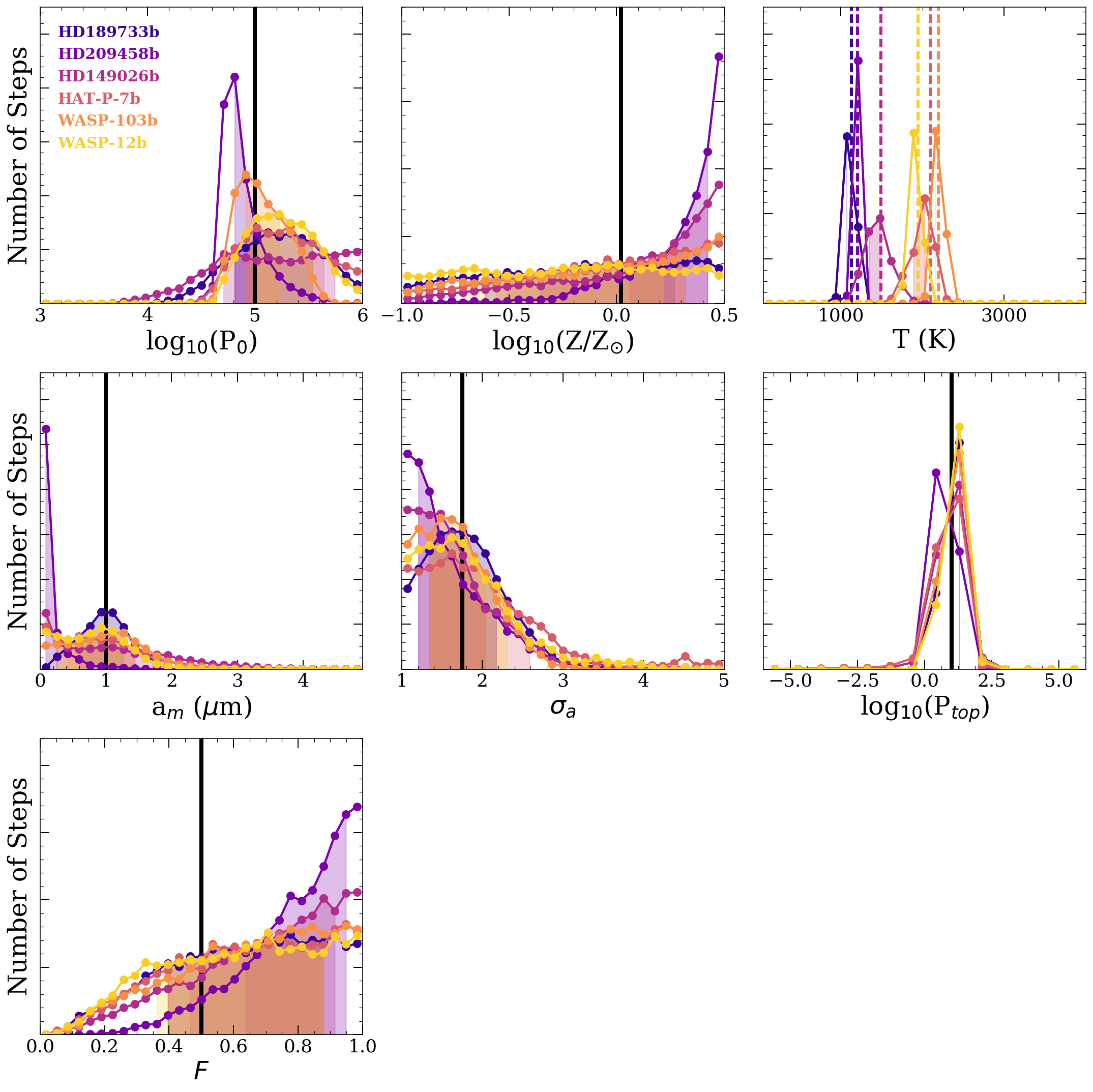}
    \caption{The posteriors for the hazy uniform-temperature control case. Colored dots and lines correspond to histograms of the MCMC chain for different objects. The shaded regions cover the range which falls between the sixteenth and eighty-fourth percentile. Vertical lines mark the true values of parameters used to simulate the data. Note that the units for P$_0$ and P$_{top}$ are both bars.}
    \label{fig:hazy_hists_uniform}
\end{figure}

\begin{figure}
    \centering
    \includegraphics[width=\textwidth]{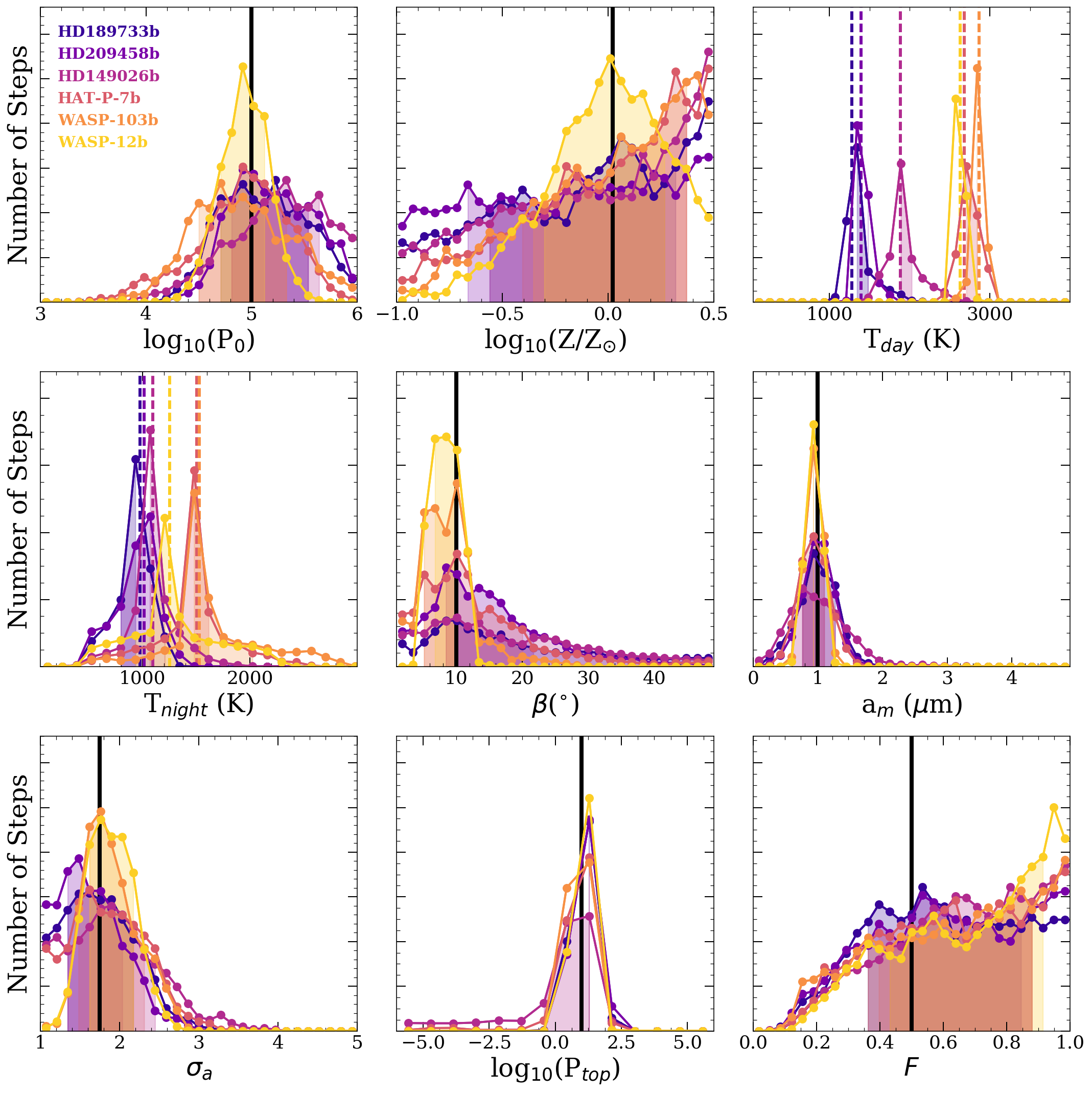}
    \caption{Posteriors for the hazy control case with a day-night temperature gradient. Colored lines and dots are histograms of the MCMC chains for each object. Shaded regions cover the sixteenth through eighty-fourth percentile. Vertical lines mark the true values of parameters used to simulate data.}
    \label{fig:hazy_hists_tgrad}
\end{figure}

\begin{figure}
    \centering
    \includegraphics[width=\textwidth]{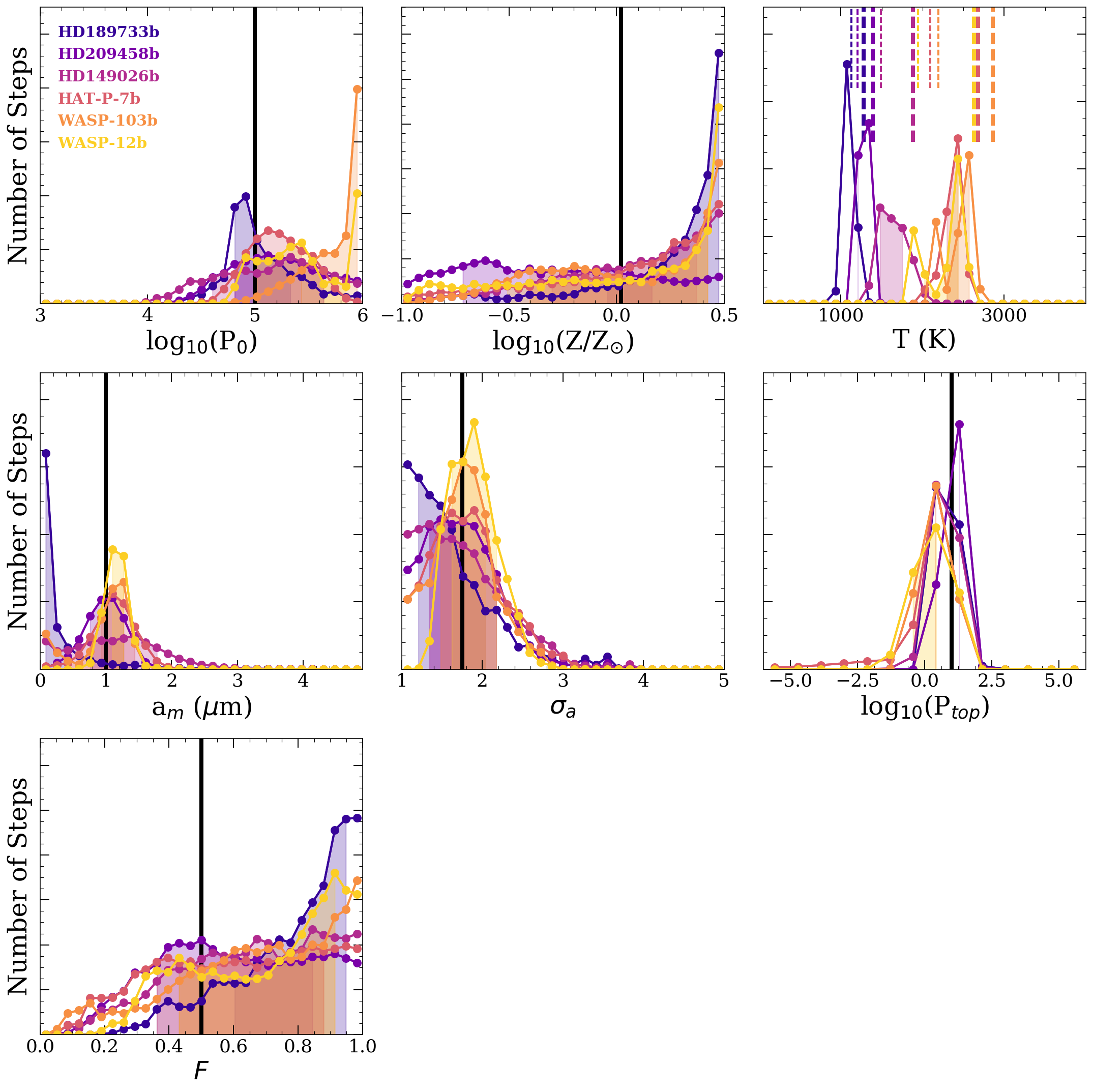}
    \caption{The posteriors for the hazy test case. Colored lines and dots are histograms of the MCMC chains for each object. Shaded regions indicate the span of the sixteenth through eighty-fourth percentile. Vertical lines mark the true values used to generate the data. There is not a single well-defined temperature, so we use shorter thinner lines to mark the average between the day-side and the night-side temperatures, and longer thicker lines to mark the day-side temperature.}
    \label{fig:hazy_hists_mixed}
\end{figure}

Figure \ref{fig:hazy_bestfits} shows the simulated data (colored dots with shaded error envelope) and spectra corresponding to the median values of the posteriors (black dashed lines). We again show the uniform temperature and temperature gradient control cases and the mixed fit test case. One can see that the haze has filled in a lot of the troughs in gaseous absorption. For the hottest planets with a temperature gradient, the resonance feature for MgSiO$_3$ is very prominent around 10 $\mu$m. All the best-fit spectra seem to agree well with the simulated data except for the mixed fits for WASP-103b and WASP-12b, the two objects with the hottest day-side temperatures and the largest difference between their day-side and their night-side temperature. The single temperature model has trouble fitting both the aerosol and the top of the CO absorption at 4.5 $\mu$m. The posteriors for all the parameters in the hazy uniform-temperature control case and the hazy temperature gradient control case are shown in Figures \ref{fig:hazy_hists_tgrad} and \ref{fig:hazy_hists_uniform} respectively. The posteriors for all the parameters in the hazy test case are shown in Figure \ref{fig:hazy_hists_mixed}. 

For the hazy uniform-temperature control case (Figure \ref{fig:hazy_hists_uniform}), the temperatures are still accurate, though some precision is lost relative to the clear atmosphere. The reference pressure, P$_0$, has a looser constraint than the clear atmospheres as well. It even becomes a lower limit for some objects, but it is still consistent with the true value. The cloud-top pressure is precise and accurate for all the objects. $F$ is always just a lower limit. The metallicity has a flat posterior (HD189733b, HAT-P-7b, WASP-103b, and WASP-12b) or a lower bound within our priors (HD209458b and HD149026b). The poor constraint on metallicity is due to a strong degeneracy between the fraction of available material incorporated into the haze ($F$), the modal particle size (a$_m$), and the metallicity ($Z$). The reference pressure (P$_{0}$) is also degenerate with metallicity when aerosols obscure gas features in the optical wavelength range. The modal particle size and size-dispersion are retrieved for HD189733b, but their posteriors only provide upper limits for HD209458b, HD149026b, HAT-P-7b, WASP-103b, and WASP-12b. 

The results for the hazy temperature gradient control case (Figure \ref{fig:hazy_hists_tgrad}) show that, rather than heightening degeneracies, the presence of both a day-night temperature gradient and a thick haze provided tighter constraints on P$_0$ and $Z$, than a thick haze in a uniform atmosphere, and provided tighter constraints on temperatures and $\beta$ than a clear atmosphere with a temperature gradient. The presence of the haze tightens the constraints on both day- and night-side temperatures significantly because it enables a much better constraint on $\beta$. Recall that, in \S \ref{sec:aerosol_paramstudies}, we saw that the presence of a slab aerosol at altitude tends to steepen the dependence of the transit spectrum on $\beta$. Similar to the hazy uniform-temperature control case, the posteriors for P$_0$ are broader than when clear, and they have a tail towards higher values. However the effect is not so severe as for the uniform-temperature hazy spectra. The posteriors for $F$ still only provide a lower limit, but the other aerosol properties are all accurately and relatively precisely retrieved. For the day-night temperature gradient control case, we get a flat posterior or a lower bound within the posteriors for all the objects except WASP-12b. The presence of a day-night temperature broke degeneracies between $Z$ and P$_{0}$ for hazy WASP-12b, allowing us to retrieve a relatively precise and accurate metallicity measurement, even in the presence of a thick haze. However, it is not as precise as the measurement for a clear atmosphere. WASP-12b has both the highest day-side temperature and also a higher SNR than the next hottest object WASP-103b, so it has the tightest constraint on $\beta$ and P$_{0}$ which allows a tighter constraint on $Z$. For the other objects, there is only a weak lower limit on metallicity, but the presence of a thick haze can weaken or erase constraints the metallicity in many planets, regardless of whether they exhibit day-night temperature gradients.

In the test case with a mixed fit (Figure \ref{fig:hazy_hists_mixed}), we see that, even when a haze is present and spans the full limb, one will obtain biased results if they ignore day-night temperature gradients. The MCMC fits find much broader posteriors for the temperature than for the clear results. The fits for WASP-103b and WASP-12b are actually multi-modal with one peak at the true terminator temperature and one peak skewed towards the day-side temperature. The reference pressure formally agrees with the true value for all the objects, but one of the modes for WASP-103b and WASP-12b pushes up against the deepest allowed reference pressure. None of the posteriors formally disagree with the true metallicity, but we do see different results than the control case with uniform temperatures and the control case with a temperature gradient. A wider prior on metallicity would be better to fully tease out the implications, but we can see a few hints from the posteriors. WASP-12b and HD189733b take on the shape of a lower-bound that is tending towards excluding the true metallicity. HD209458b has shifted to an almost uniform posterior across the allowed metallicities, whereas for the fit to a truly uniform temperature atmosphere it was possible to rule out the lowest metallicities. HAT-P-7b and HD149026b are largely unaffected. The recovery of the aerosol size distribution performs roughly as well as or better than the uniform-temperature hazy control case. This speaks to the nature of the aerosol spectral signatures imprinted in our simulated data. When the data is generated using a day-night temperature gradient for the hotter objects the wavelength-dependent cross section of MgSiO$_3$ is clearly imprinted on the spectra, but when the data was generated using a single temperature the haze imparts a predominantly gray opacity on the spectrum. The posteriors for $F$ still show a loose lower bound for all the objects. However, the posterior for  HD189733b formally excludes the true value of $F$. This was the object that seemed to be adequately fit by a single T-P profile when it was clear, due to it's small difference between day-side and night-side temperature. Adding in a haze has made it possible and necessary to use a model with a temperature gradient instead.

\subsection{Cloudy Atmospheres}\label{sec:cloudy_fits}

\begin{figure}
    \centering
    \includegraphics[width=\textwidth]{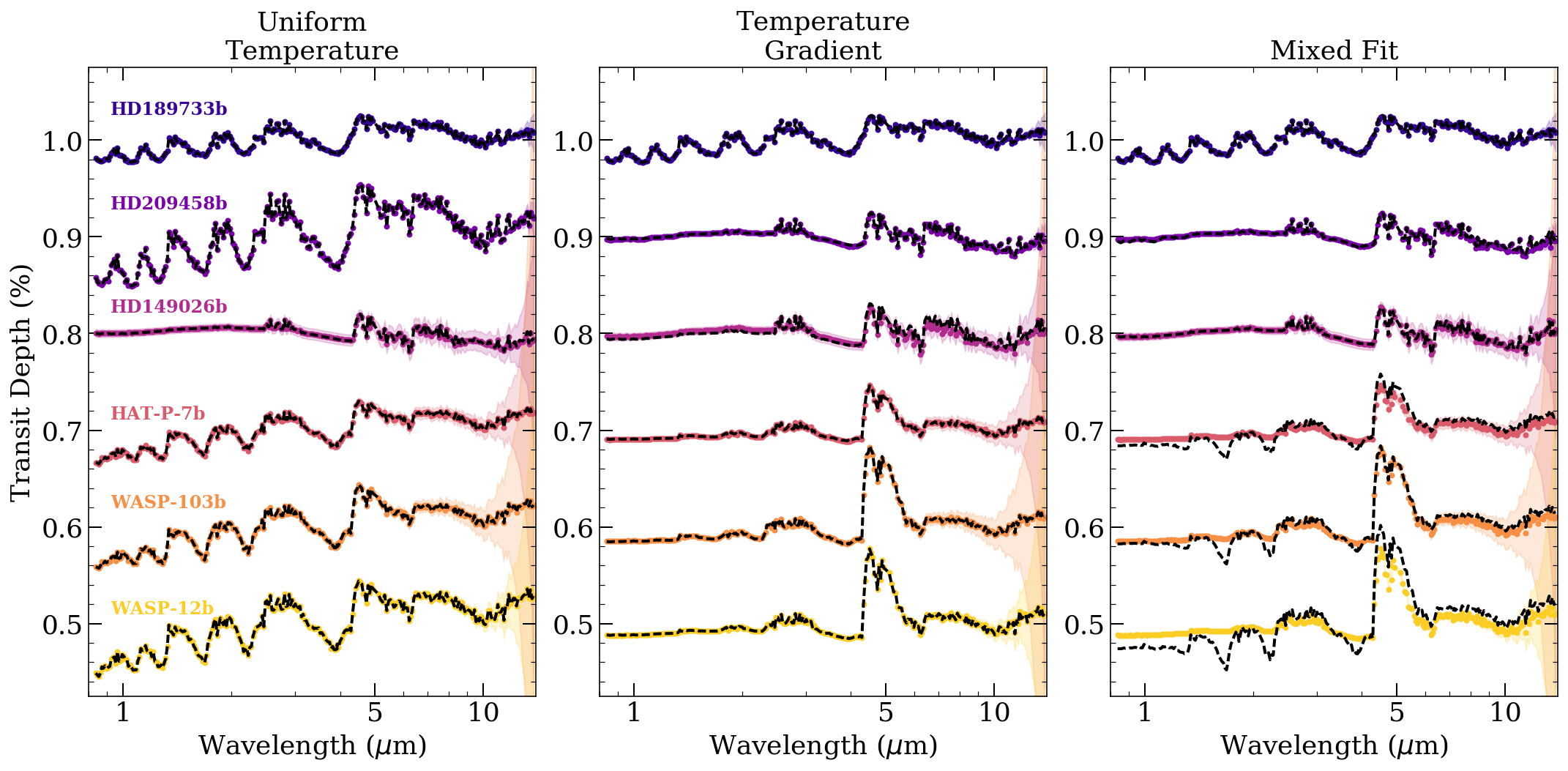}
    \caption{Cloudy simulated data (colored dots with error envelopes) and fits (black dashed lines) corresponding to the median parameter values of the MCMC chains. All spectra were modelled with an equilibrium cloud of MgSiO$_3$ that tapered of with $\alpha$=2. The log-normal particle size distribution had a modal size of 1 $\mu$m and a dispersion of 1.75. Parameters for the different objects are taken from Table \ref{tab:cowan_objects}. When data was simulated using a single temperature (left panel), the average between the day-side and night-side temperature was used. This meant that only HD149026b formed a significant enstatite cloud. HD189733b does not form a significant cloud, even when the day-night temperature gradient model is used to simulate data. See Figure \ref{fig:cclines} for the condensation curve for MgSiO$_3$ used in our model.}
    \label{fig:cloudy_bestfits}
\end{figure}

Some evidence hints that clouds may be condensing on the night sides of hot Jupiters and evaporating on the day sides, so we decided to include an experiment that approximates this behavior. We simulated transit spectra which include a cloud of MgSiO$_3$ according to the phase equilibrium cloud model described in \S \ref{sec:aerosols}. Figure \ref{fig:cloudy_bestfits} shows the simulated data and best-fit models. For the data simulated with a uniform temperature atmosphere, it should be noted that only HD149026b had a temperature that would allow MgSiO$_3$ to condense and contribute significantly to the transit spectrum (see left panel). When the planets had a day-night temperature gradient, MgSiO$_3$ could condense somewhere within the limb for all the objects except HD189733b (see center and right panels). The mixed fits with an equilibrium cloud for the cooler objects (HD149026b, HD209458b, and HD189733b) would pass a goodness-of-fit criterion, but the hotter objects (HAT-P-7b, WASP-103b, and WASP-12b) would certainly fail. We examine the posteriors of these retrievals in more detail below.

\begin{figure}
    \centering
    \includegraphics[width=\textwidth]{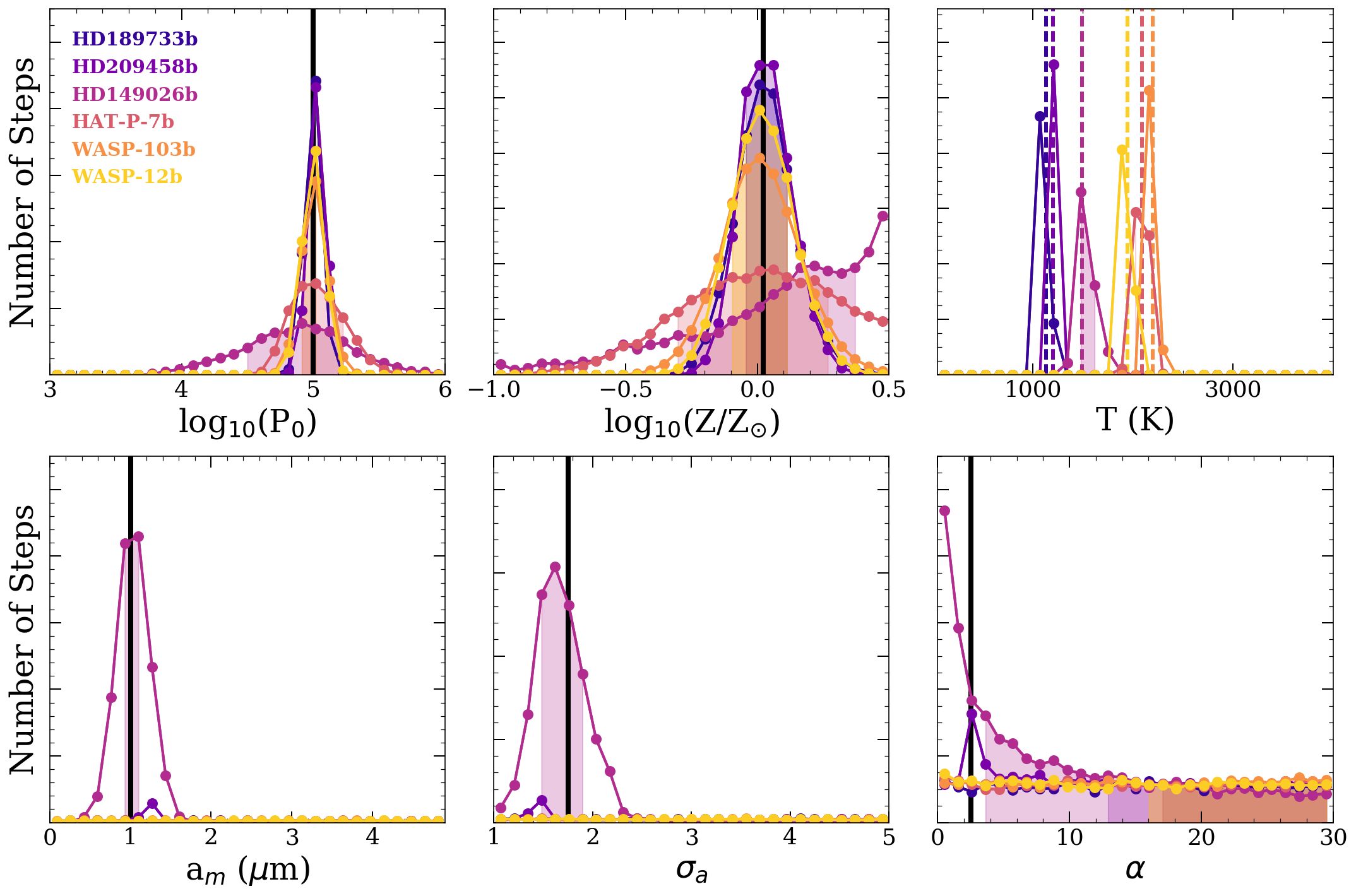}
    \caption{The posteriors for the cloudy uniform-temperature control case. Colored lines and dots show histograms of the MCMC chains for each object. Shaded regions mark the span between the sixteenth and eighty-fourth percentile. Vertical lines denote the true values of parameters used to simulate the data.}
    \label{fig:cloudy_hists_uniform}
\end{figure}

\begin{figure}
    \centering
    \includegraphics[width=\textwidth]{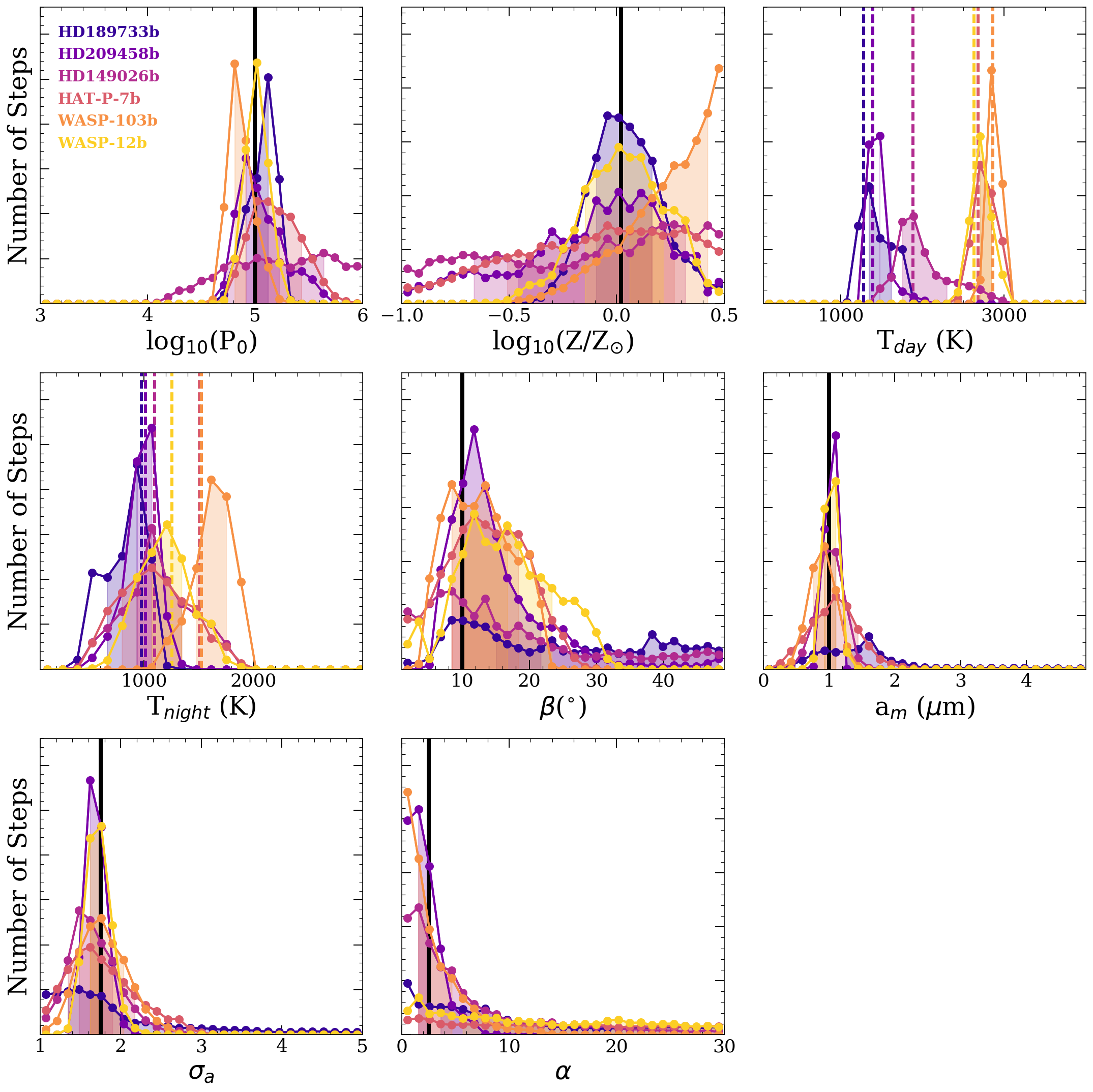}
    \caption{The posteriors for the cloudy control case with a day-night temperature gradient. Colored lines and dots show histograms of the MCMC chains for each object. Shaded regions mark the span between the sixteenth and eighty-fourth percentile. Vertical lines denote the true values of parameters used to simulate the data.}
    \label{fig:cloudy_hists_tgrad}
\end{figure}

\begin{figure}
    \centering
    \includegraphics[width=\textwidth]{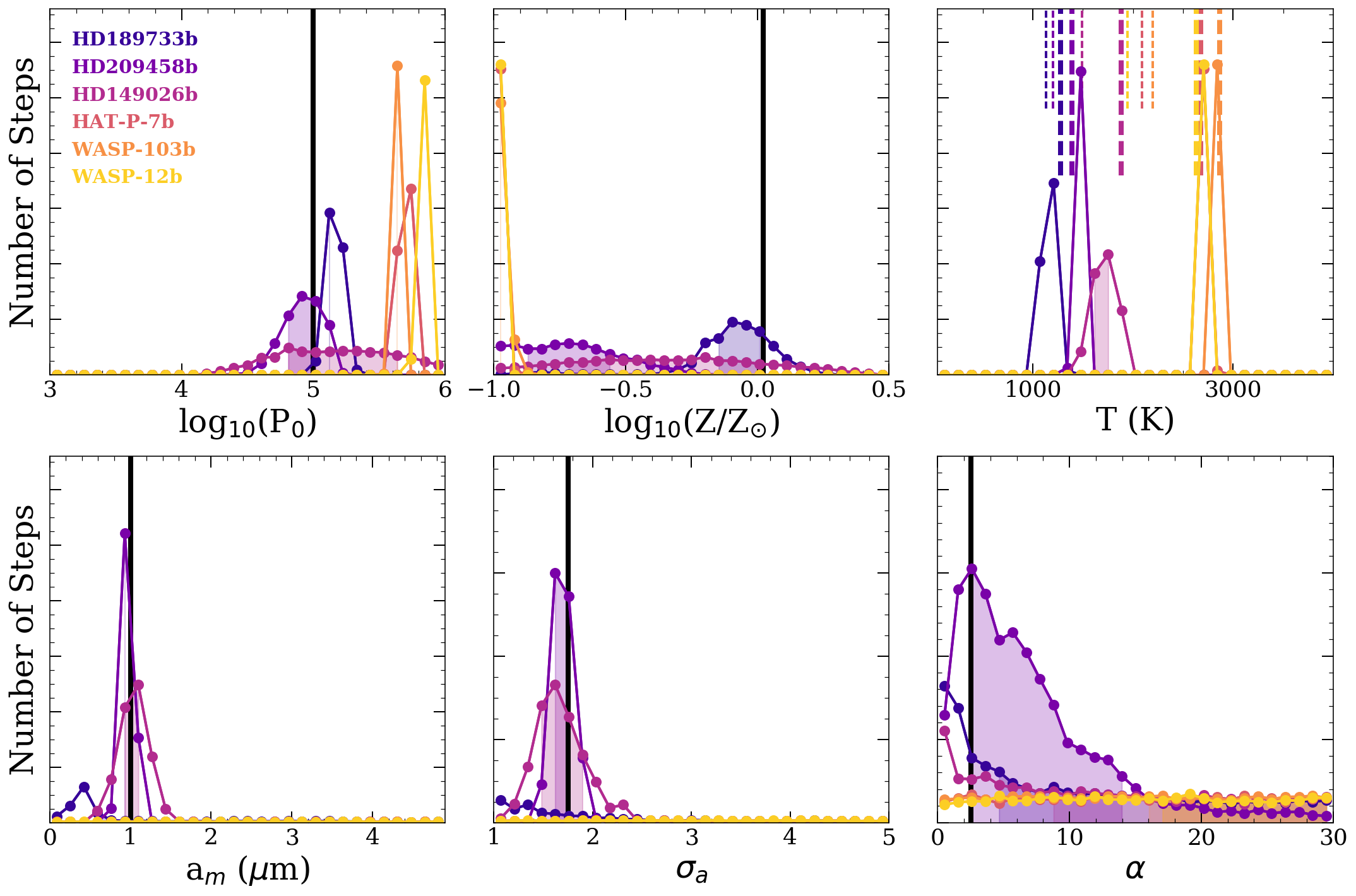}
    \caption{The posteriors for the cloudy test case. Colored lines and dots show histograms of the MCMC chains for each object. Shaded regions mark the span between the sixteenth and eighty-fourth percentile. Vertical lines denote the true values of parameters used to simulate the data. There is not a single well-defined temperature for the mixed fit, so we use shorter thinner lines to mark the average between the day-side and the night-side temperatures, and longer thicker lines to mark the day-side temperature.}
    \label{fig:cloudy_hists_mixed}
\end{figure}

Figure \ref{fig:cloudy_hists_uniform} shows all the posteriors for the control case with a uniform temperature. Since only HD149026b can form a significant enstatite cloud, that is the only posterior which places a constraint on the particle size distribution (maroon lines and dots). The retrieved values for modal particle size and size dispersion are both accurate and precise. The posterior for $\alpha$ only provides a very week upper bound, implying that the majority of the cloud opacity is contributed near the cloud base pressure, so varying $\alpha$ has little effect on the shape of the transit spectrum. The reference pressure is still accurate but less precise than for a clear atmosphere, due to the degeneracy with $Z$. The retrieval for HD149026b can only place a lower limit on metallicity (within our priors) due to a degeneracy with $\alpha$ and P$_{0}$. HD209458b shows a slight hint of a cloud (lighter purple lines and dots). Posteriors for all the other objects have essentially flat posteriors for a$_m$, $\sigma _a$, and $\alpha$, then similar posteriors to the clear atmospheres for P$_0$, $Z$, and $T$. Such a result from your fit would indicate that you should run it again leaving the cloud parameters out.

Figure \ref{fig:cloudy_hists_tgrad} shows all the posteriors for the cloudy control case simulated and fit with a temperature gradient. It seems that, if condensed clouds are present and well-described by equilibrium models, then a lot can be learned about both the cloud particle size distribution and the day-night temperature structure. Similar to the results for the slab aerosol, the presence of a cloud improves the constraint on $\beta$ and thus on the day- and night-side temperatures relative to the clear case. This comes at the cost of broadening the constraint on metallicity relative to the clear case, but, for four of the objects at least, the posterior still shows a distinct peak centered on the true metallicity. For HD189733b, HD209458b, and HD149026b, the constraint on metallicity is actually more accurate than the clear case with a temperature gradient, though it is also much less precise. The modal particle size and the dispersion of the log-normal size distribution are well constrained for all the equilibrium cloud fits, but $\alpha$ is just an upper bound in every case. Rather than confusing the picture, it seems that the combined presence of condensing clouds and temperature gradients can provide deeper insight into the atmospheres of tidally-locked planets.  

Figure \ref{fig:cloudy_hists_mixed} shows the results for the test case, using a single temperature to fit data simulated with a temperature gradient and an equilibrium cloud. Most notably, in the three planets with the hottest day sides, a single temperature cannot account for both the tall CO feature around 5 $\mu$m and the formation of an enstatite cloud within our equilibrium cloud parameterization. The MCMC model ends up favoring a higher temperature, so there is no constraint on particle size, width of distribution, or the relative scale height of gas and aerosol for HAT-P-7b, WASP-103b, and WASP-12b. This can be seen in the right-hand panel of Figure \ref{fig:cloudy_bestfits}. The black dashed lines overlying the yellow, orange, and pink data don't have any clouds in them. WASP-12b, WASP-103b, and HAT-P-7b almost exactly retrieve their day-side temperature alone. In order to bring the gaseous absorption in the optical up to near the depths of the transit spectra with a cloud, the reference pressures are biased to higher values. Meanwhile, to bring the absorption in the mid infrared down and flatten things out, the metallicities are biased to lower values. The mixed fits for HD209458b and HD149026b are able to replicate the data very well. To form the cloud base at the right pressure level, HD209458b retrieves a temperature even higher than its day-side temperature, while  HD149026b retrieves a temperature just below its day-side temperature. The correct reference pressure, modal particle size, and dispersion of the particle size distribution are constrained for these objects, but the metallicities skew towards values that are too low to get approximately the right balance between gaseous opacity and cloud opacity in the transit spectrum. For HD189733b, there is not much cloud at all, so the posteriors for P$_0$, $Z$, and $T$ are similar to the clear mixed fits. Some very small amount of cloud must be present though, since posteriors for a$_m$, $\sigma _a$, and $\alpha$ are not simply flat lines both here and in the temperature gradient control case. The posterior for a$_m$ is mostly flat, but it has a peak that is much smaller than the true a$_m$, whereas the posterior for a$_m$ peaks at the correct value in the cloudy temperature gradient control case. Both the structural parameters and the aerosol-related parameters end up biased in this case. 

These results show that if the aerosols in tidally-locked hot Jupiters are condensing species, then ignoring day-night temperature gradients in models can lead to severe and complicated biases. For planets with extremely hot day-sides, the inadequacy of the single temperature model will be apparent, but for others it won't be.  When the day-night temperature gradient is accounted for, one obtains accurate constraints on structural and aerosol-related parameters in most cases. 

Taken all together, the results of these experiments show that the importance of day-night temperature gradients is highly dependent on the properties of the object in question, the $SNR$ of the data, and the presence or absence of aerosols. When HD189733b contains a haze, it is important to account for the day-night temperature gradient, but, when it is clear, the day-night temperature gradient can almost be ignored. This planet has the smallest difference between its day-side temperature and its night-side temperature. HD209458b has a similar difference between its day-side temperature and night-side temperature, but a larger $SNR$, especially when the atmosphere is clear. If there is a haze, then fitting HD209458b with a single T-P profile does not retrieve significantly biased results for any parameter, but, if it is clear, then a single T-P profile biases metallicity to the point that it formally disagrees with the true value. If it has a cloudy atmosphere, then the correct aerosol properties are retrieved, but the wrong metallicity and temperature. It is always important to account for the day-night temperature gradient in WASP-103b and WASP-12b whether clear, hazy, or cloudy. These two planets have the largest difference between their day-side temperature and their night-side temperature. If condensed clouds are present in the atmospheres of WASP-103b and WASP-12b, it may be apparent that a single T-P profile model is inadequate. HAT-P-7b and HD149026b have large differences between their day- and night-sides, but also have lower $SNR$, and thus, larger uncertainties in their retrieved parameters. For these two planets, the biases induced by ignoring temperature gradients didn't shift retrieved metallicities to the extent that they formally disagreed with the true values. The exception is if HAT-P-7b contains a condensed cloud. In that case, the best fit using a single T-P profile retrieved an upper bound on the metallicity that excluded the true value, but the corresponding spectra would likely fail to satisfy goodness-of-fit criteria. In the cases investigated here, the effects of aerosols and the effects of day-night temperature gradients are not degenerate. In fact, if both are present in the atmosphere and accounted for in the retrieval model, their combined effects can actually tighten constraints on relevant parameters.

\section{Summary and Conclusions} \label{sec:conclusions}

We began this paper by presenting METIS, a new code capable of computing transit spectra for an arbitrary 3D temperature-pressure structure and capable of performing Bayesian parameter estimation in some simplified cases (a parameterized day-night temperature gradient with azimuthal symmetry or a single 1D temperature pressure structure that describes the whole terminator region). This tool is thus well-suited to modeling transit spectra for planets with variation about their limbs.

Puffy hot planets and planets with shorter periods make it easier to obtain high quality transit spectra. Many of these planets are tidally locked and thus exhibit large day-night temperature differences and other inhomogeneities about their limbs. This means that many top-quality transit spectra candidates will be tidally locked. If structures inferred from GCM modeling and phase curve observations are correct, the effects of day-night temperature gradients cannot be ignored in transit spectra retrievals. It has been shown previously that temperature gradients from day to night effect transit spectra retrievals in some cases, tending to shift them towards the higher day-side temperatures, which then biases the retrieved chemical abundances \citep{Caldas2019}. The first relevant retrieval experiments were done for clear atmospheres with varying temperatures, but uniform chemistry. \citealt{Pluriel2020} expanded on this to consider atmospheres which vary the H$^{-}$ abundances and corresponding opacity between the day and night side. 

We expanded this line of study by varying the abundances of all significant species between the day and night side and incorporating different types of aerosols. For some of these planets, the day-side temperatures seem too hot for aerosols to form, but species could condense on the night side. In other cases, aerosols originating from photochemistry or condensation could be present throughout the limb of the planet. We used METIS to explore the effects of aerosols and day-night temperature gradients on transit spectra, focusing in particular on retrievals for transit spectra simulated with the spectral resolution, SNR, and wavelength coverage that are expected for JWST. 

We will now enumerate the questions addressed in this paper and summarize the answers implied by our results.
\begin{enumerate}
    \item \textbf{Do clouds and hazes make it more or less important to account for the presence of a day-night temperature gradient when interpreting transit spectra?} We find that the presence of aerosols at altitude tends to increase the importance of accounting for day-night temperature gradients when interpreting transit spectroscopy of tidally-locked exoplanets. This is particularly true for clouds formed through condensation rather than hazes formed through photochemical processes because the inferred temperature consistent with gaseous absorption features arising from the day-side may be inconsistent with a cooler temperature needed for the aerosol to condense on the night-side. Luckily, it seems likely that if Nature presents us with a case like this, we will be able to tell that models with a single T-P profile are inadequate given reasonable JWST-like data. Another reason that aerosols expand the importance of accounting for day-night temperature gradients is that, when aerosols are present, the transit spectrum tends to be shaped at higher altitudes, where it is more sensitive to variations in the day-night structure. In this case, the magnitude of the day-night difference doesn't need to be as large for the effect to be important. This also means that, when one does use a day-night temperature gradient in the retrieval, the parameters for T$_{day}$, T$_{night}$, and $\beta$ are much better constrained when aerosols are present than for a purely clear atmosphere, although the metallicity is often more poorly constrained.
    \item \textbf{For which objects will these effects be significant? Can we retrieve reliable information from the transit spectra of tidally locked exoplanets? } We do not provide an exhaustive or definitive answer to these questions, but the results shown are suggestive. Our toy model is able to retrieve meaningful constraints on day- and night-side temperatures, metallicity, and the width of the day-night transition for WASP-12b, WASP-103b, and HD209458b. HD189733b is fairly well described by a single T-P profile. HAT-P-7b and HD149026b may not have sufficient $SNR$ to make the more complex day-night temperature gradient model better than a single T-P profile model, unless condensed clouds are present on the night-side of HAT-P-7b. However, this model is more suitable for testing whether day-night temperature gradients will have significant effects on transit spectra than for performing retrievals on actual data. It assumes an isothermal T-P for each longitude, full chemical equilibrium, and a solar C/O ratio. In reality those assumptions may not hold, and a true retrieval must be flexible enough to let the data drive the conclusions. There may also be further variation between the poles, and the morning and evening terminators that must be accounted for. 
    \item \textbf{What are the biases on retrieved atmospheric parameters from ignoring day-night gradients?} With the addition of fully varying chemistry across the day-night transition, we find that the retrieved metallicities for our six planets sometimes scatter above the true metallicity and sometimes scatter below the true metallicity, when the atmospheres are clear. When the atmospheres have a thick enstatite haze, then the metallicity is degenerate with the fraction of material incorporated into the aerosol, regardless of whether we fit the data with the correct type of temperature structure or not. This degeneracy is a bit weaker for the day-night temperature gradient compared to the uniform temperature atmosphere. When the atmospheres have an enstatite cloud that forms only where temperatures and pressures are favorable, then the metallicity is better constrained than the slab-aerosol case.
\end{enumerate}

With the advent of JWST and ARIEL, researchers will need to determine how to interpret high-$SNR$ transit spectra with broad wavelength coverage for objects that vary significantly between the day- and night-sides of their limbs. At the very least we must avoid biased results, and ideally we will extract extra information about the transition from day to night. This work represents a first step in this direction. It is important to continue investigating as-yet overlooked factors and to develop approaches for mitigating already identified pitfalls as we prepare to interpret the exoplanet transit spectra that will become available in the near future. 

\acknowledgments
The authors would like to acknowledge support for this research under NASA 
WFIRST-SIT award \# NNG16PJ24C and NASA Grant NNX15AE19G. This research has made use of the NASA Exoplanet Archive, which is operated by the California Institute of Technology, under contract with the National Aeronautics and Space Administration under the Exoplanet Exploration Program.

\vspace{5mm}
\software{astropy \citep{astropy}, numpy (\citealt{numpy1}; \citealt{numpy2}), scipy (\citealt{scipy1}; \citealt{scipy2}), matplotlib \citep{matplotlib}, emcee (\citealt{emcee1}; \citealt{Goodman2010}), corner (\citealt{corner}), mpi4py (\citealt{Dalcin2011})}

\appendix
\section{Approximating JWST-like Measurements} \label{sec:noise_model}
We used the online interface of the community tool Pandexo\footnote{https://exoctk.stsci.edu/pandexo/calculation/new} and the reccomendations of \citealt{Greene2016} to simulate JWST-like observations for the planets in Table \ref{tab:cowan_objects}. Pandexo can calculate the precision on the depth measurement of a single transit with the four instruments on JWST in a variety of modes (NIRISS, NIRCam, NIRSpec, and MIRI). As input, Pandexo requires a stellar SED model, the apparent magnitude of the star, the planet's spectrum (primary or secondary), the transit duration, a fraction of time spent observing in-transit versus out-of-transit, the number of transits, an exposure level considered to be the saturation of the detector, and an optional user-defined noise floor. For our calculations, we just use the pre-loaded target properties and default Phoenix spectral models (taken from Phoenix Stellar Atlas (Husser et al. 2013)), set the transit radius to a constant value with wavelength, set the saturation to 80\% of full-well, and allow Pandexo to optimize the number of groups per integration. 

\begin{figure}
    \centering
    \includegraphics[width=\textwidth]{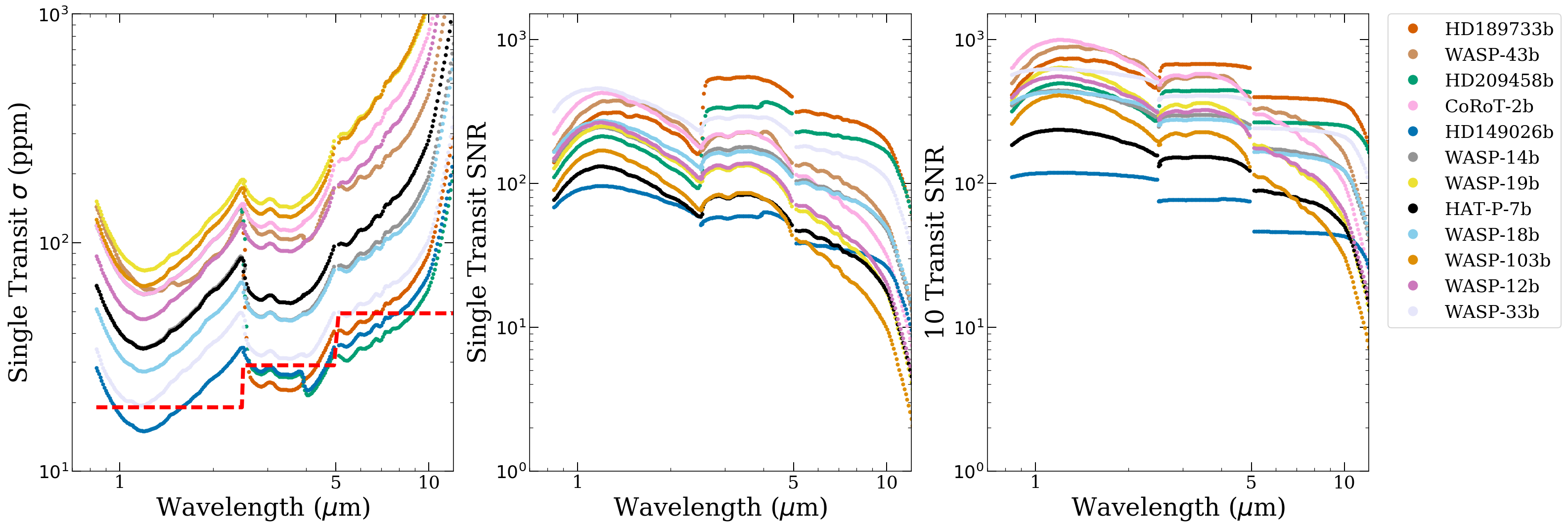}
    \caption{Signal-to-noise calculations for JWST observations of the objects in Table \ref{tab:cowan_objects} based on Pandexo results for a single transit observation with NIRISS-SOSS, NIRCam, and MIRI-LRS and a noise floor adopted is from \citealt{Greene2016}. The left panel shows the sing-transit precision on the depth measurement and the noise floor (red dashed line). The center panel shows the $SNR$ for a single transit with each instrument. The right panel shows the $SNR$ for 10 transits with each instrument.}
    \label{fig:noise}
\end{figure}

\citealt{Greene2016} focus on NIRISS, NIRCam and MIRI, and suggest re-binning the native resolution of the instruments to R$\sim$100 as a compromise between spectral information and adequate $SNR$, so we used Pandexo to compute the precision in these scenarios without imposing a noise floor (see colored dots in the left panel of Figure \ref{fig:noise}). These single-transit error bars (N$_{single}$) can be scaled to replicate the result of stacking multiple transits by dividing by the square root of the number of transits observed ($\sqrt{n_{tr}}$). One must also impose a noise floor to reflect the reality that instrument systematics and astrophysical systematics will eventually impose a noise limit that cannot be overcome. We use a noise floor in line with systematics of HST and Spitzer as recommended by \citealt{Greene2016}: 20 ppm for NIRISS, 30 ppm for NIRCAM, and 60 ppm for MIRI. This is shown as a dashed red line in the left panel of Figure \ref{fig:noise}.

We add the noise floor in quadrature with the scaled Pandexo noise calculation to get the following expression for SNR:
\begin{equation}
    SNR(\lambda) = \frac{\frac{\delta F}{F}(\lambda)}{\sqrt{\frac{1}{n_{tr}}N_{single}(\lambda)^2+N_{floor}(\lambda)^2}} ~~,
\end{equation}
where $\frac{\delta F}{F}(\lambda)$ is the fractional change in brightness or the transit depth equal to ($R_P$/$R_*$)$^{2}$, $N_{single}$ is the error computed by Pandexo, $n_{tr}$ is the number of transit observations to be stacked, and $N_{floor}$ is the noise floor from \citealt{Greene2016}. When we simulate data we simply make random draws from normal distributions about the true depths with widths $\sqrt{\frac{1}{n_{tr}}N_{single}(\lambda)^2+N_{floor}(\lambda)^2}$, and assign the error bars for each data point as $\sqrt{\frac{1}{n_{tr}}N_{single}(\lambda)^2+N_{floor}(\lambda)^2}$.

The center panel of Figure \ref{fig:noise} shows the $SNR$ for a single transit with each instrument/mode (that is 4 total transits, since we use NIRISS, two modes of NIRCam, and MIRI to cover the full wavelength range), and the right-most panel shows the $SNR$ for 10 transits with each instrument/mode (that is 40 total transits). In this 10-transit case some objects have begun to asymptote to the noise floor inferred based on HST and Spitzer performance. This shows up as the curves turning to a flat step function, like the dark blue dots for HD149026b. An object's maximum possible $SNR$ thus depends on how the noise floor compares to its (R$_P$/R$_*$)$^{2}$.

This formulation of $SNR$ is the convenient one for simulating observations. Of course, the real ``signal" in transit spectroscopy also involves the scale height of the atmosphere. In terms of information encoded in a transit spectrum, one would more naturally form a definition of $SNR$ which quantifies how the variation of transit depth with wavelength compares to the precision of the depth measurement at each wavelength. Since our purposes here are to simulate realistic JWST-like observations of 12 specific objects, we stick with the $SNR$ as depth divided by the precision of the depth measurement.   

\section{Markov Chain Monte-Carlo Retrievals}\label{sec:mcmc_methods}
We use the python package {\tt emcee} to do MCMC retrievals with METIS. When running the chains, we adjust the burn-in time as needed, but always obtain a final chain length of 140,000 steps. We assume uniform priors on all of the model parameters which are summarized in Table \ref{tab:priors}. For all the retrieval experiments in this work, we simulated data as if observations were done with JWST's NIRSS, NIRCam mode I, NIRCam mode II, and MIRI for 10 transits each, then binned to a spectral resolution of 100. This means that we have ~300 wavelengths. We use a moderately sized grid of 100 altitudes, 70 longitudes, and 70 latitudes in the METIS forward models. Under these conditions, the chains take somewhere between 5 hours and 30 hours to complete depending on whether it is a clear or cloudy model, and whether there is a day-night temperature gradient or a uniform temperature atmosphere. In all cases, we assume that the mass of the planet, the reference radius of the planet corresponding to P$_0$, a solar C/O ratio, and the radius of the star are known and that stellar surface inhomogeneities are not contributing \citep{Rackham2017}. 

\begin{table}[]
    \centering
\begin{tabular}{c|c|l}
     Parameter& Prior &Description\\
     \hline
     P$_0$& 0.01 bars $<$ P$_0$ $<$ 10 bars & Reference pressure corresponding to known radius.\\
     Z & 0.1 $<$ Z/Z$_{\odot}$ $<$ 3.16 & Bulk metallicity \\
     T$_{day}$& 50 K $<$ T$_{day}$ $<$ 4000 K & Temperature or day-side temperature\\
     T$_{night}$& 50 $<$ T$_{night}$ $<$ T$_{day}$ & Night-side temperature\\
    $\beta$ & 1$^{\circ}$ $<$ $\beta$ $<$ 179 $^{\circ}$ & Angular width of day-night transition \\
    a$_m$& 0.001 $\mu$m $<$ a$_m$ $<$ 100.0 $\mu$m &Modal particle size\\
    $\sigma _a$& 1.0 $<$ $\sigma _a$ $<$ 50.0 &Width of log-normal particle size distribution\\
    $\alpha$ & 0.001 $<$ $\alpha$ $<$ 100 & Ratio of aerosol scale height to gaseous scale height \\
    P$_{top}$&   10$^{-7}$bars $<$ P$_{top}$ $<$ P$_0$& Top pressure cut-off of aerosol\\
    F&0 $<$ $F$ $<$ 1& Fraction of available material bound up in aerosol\\
\end{tabular}
    \caption{Priors used in MCMC retrievals.}
    \label{tab:priors}
\end{table}

\section{Testing METIS Integration Scheme}\label{sec:validate}
\begin{figure}
    \centering
    \includegraphics[width=0.75\textwidth]{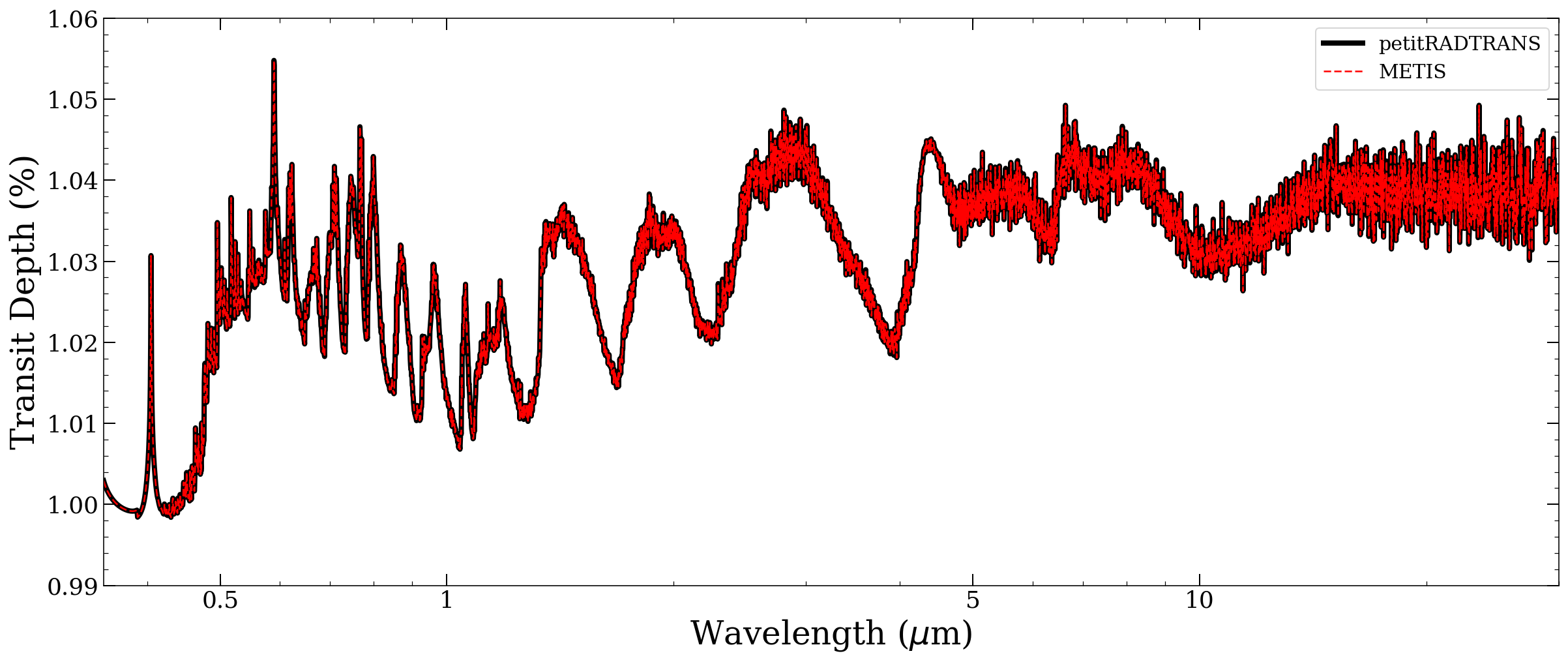}
    \includegraphics[width=0.75\textwidth]{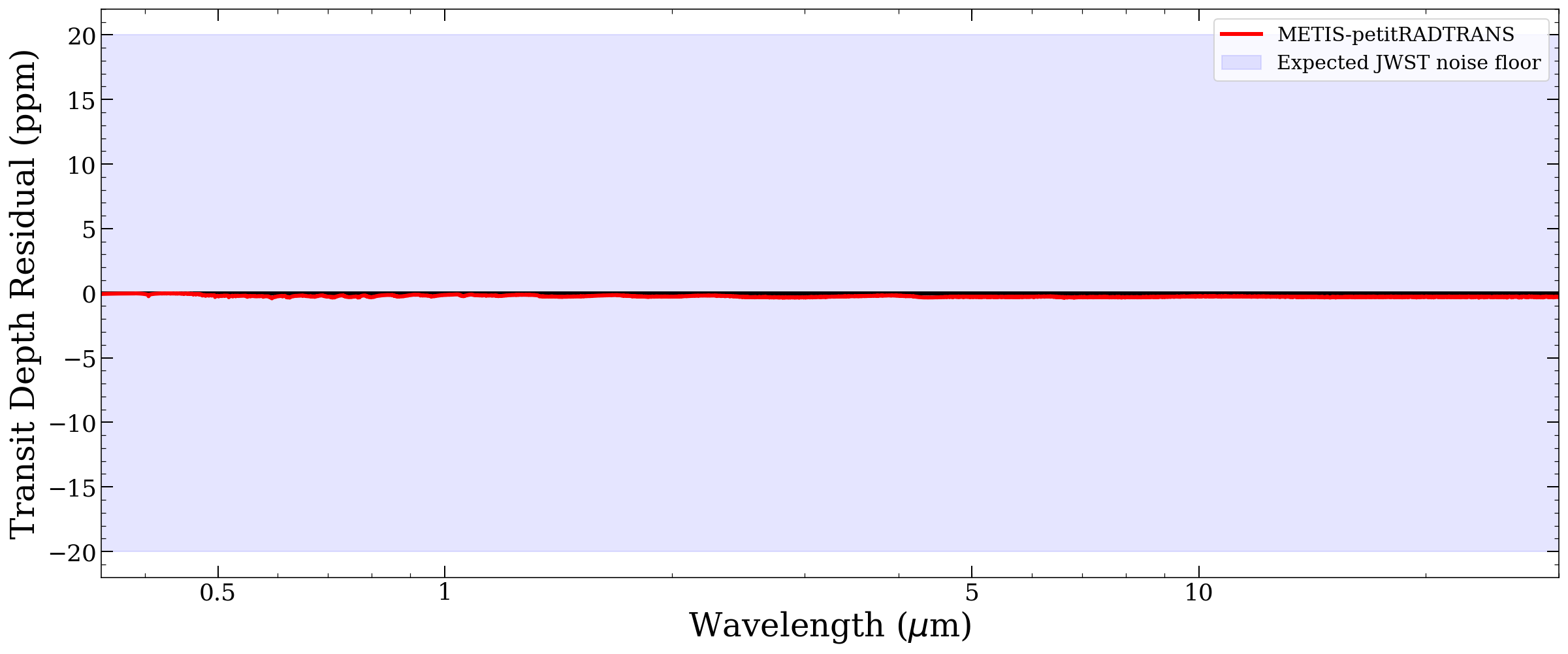}
    \caption{Comparison of METIS to petitRADTRANS. The top panel shows transit spectra in terms of the transit depth. The bottom panel shows the residuals between METIS and petitRADTRANS compared to the systematic noise floor for JWST suggested by \citealt{Greene2016}.}
    \label{fig:comparetransitcodes}
\end{figure}

We demonstrate the accuracy of our new code by comparing METIS to the peer-reviewed and publically available code petitRADTRANS\footnote{https://petitradtrans.readthedocs.io}. Figure \ref{fig:comparetransitcodes} shows transit spectra and residuals for a Jupiter-sized exoplanet with a 1500-K isothermal atmosphere and equilibrium chemistry. Continuum opacities from H$^{-}$ and CIA of H$_2$-H$_2$ and H$_2$-He, and line-by-line opacities from CH$_4$, H$_2$O, CO$_2$, CO, NH$_3$, PH$_3$, H$_2$S, TiO, VO, Na, K, and SiO are all included in the calculations.

When used the exact same opacities as petitRADTRANS in the METIS calculation to isolate the integration scheme itself. There is agreement between the two codes well below the noise floor of 20 ppm on transit \textit{depth} suggested by \citealt{Greene2016}. Note that when we use our own opacity tables we find more absorption in the infrared than petitRADTRANS and ExoTransmit. This is because we include some additional opacity sources that were not available in the petitRADTRANS library, and our tables were computed with different line lists for some species.

\end{document}